\documentclass[apj]{emulateapj}
\usepackage[sort&compress]{natbib}
\usepackage{epsf,graphics}
\usepackage{subfigure,graphicx}
\usepackage{bm}
\usepackage{mathrsfs}
\usepackage{txfonts}
\usepackage{graphicx}
\usepackage{rotating}

\begin{document}
 
\title{A Complete Spectroscopic Survey of the Milky Way satellite Segue 1: Dark 
matter content, stellar membership and binary properties from a Bayesian 
analysis} 

\author{Gregory D. Martinez, Quinn E. Minor, James Bullock, Manoj Kaplinghat}
\affil{Department of Physics and Astronomy, University of California, Irvine CA 
92697, USA}
\author{Joshua D. Simon}
\affil{Observatories of the Carnegie Institution of Washington, 813 Santa Barbara St., Pasadena CA 91101, USA}
\author{Marla Geha}
\affil{Department of Astronomy, Yale University, New Haven, CT 06520, USA}

\begin{abstract}
We introduce a comprehensive analysis of multi-epoch stellar line-of-sight 
velocities to determine the intrinsic velocity dispersion of the
ultrafaint satellites of the Milky Way. 
Our method includes a simultaneous Bayesian analysis of both membership 
probabilities and the contribution of binary orbital motion to the observed 
velocity dispersion within a 14-parameter likelihood.
We apply our method to the
Segue 1 dwarf galaxy and conclude that Segue 1 is a dark-matter-dominated 
galaxy at high probability with an {\em intrinsic} velocity
dispersion of $3.7^{+1.4}_{-1.1}$  km~s$^{-1}$.  The dark matter halo
required to produce this dispersion must have an  average density of
$\bar{\rho}_{1/2} = 2.5^{+4.1}_{-1.9} M_\odot {\rm pc}^{-3}$  
within a sphere that encloses half the galaxy's stellar luminosity.  
This is the highest measured density of dark matter in the Local Group. 
Our results show that a significant fraction of the stars in Segue 1 
may be binaries with the most
probable mean period close to 10 years, but also consistent 
with the 180 year mean period seen in the solar
vicinity at about $1\sigma$.  Despite this binary population,
the possibility that  Segue 1 is a bound star cluster with the observed
velocity dispersion arising from the orbital motion of
binary stars is disfavored
by the multi-epoch stellar velocity data at greater than 99\% C.L.    
Finally, our treatment 
yields a projected (two-dimensional) half-light radius for the stellar profile of
Segue 1  of $R_{1/2} =28^{+5}_{-4}$ pc, in excellent agreement with
photometric measurements. 
\end{abstract}


\section{Introduction}\label{sec:intro}

The discovery of faint satellites of the Milky Way has been revolutionized by 
the Sloan Digital Sky Survey (SDSS) data \citep{willman2005,zucker2006,belokurov2007}.  
These galaxies are much fainter than previously known Milky Way satellites, and 
the inferred velocity dispersions range from $\sim3$ to 8~km~s$^{-1}$
\citep{kleyna2005,martin2007,simon2007,geha2009}. Particularly at the lower end of this 
range, the inferred dispersions are susceptible to systematic biases. The most 
serious of these issues are the contribution of binary orbital motions to the 
velocity dispersion \citep{olszewski1996,hargreaves1996,odenkirchen2002,minor2010} and 
contamination of dwarf galaxy member samples by Milky Way stars 
\citep{aden2009}. These problems are most critical for the ultrafaint 
satellites with small velocity dispersions because the stellar velocity samples 
are limited in size and contributions from binary or nonmember
(Milky Way or overlapping stream) stars to the measured velocity dispersion may 
represent an appreciable fraction of the galaxy's intrinsic dispersion.  
Binaries have been the most difficult of these potential biases to correct 
because the properties of binary stars in environments beyond the solar 
neighborhood are not well known and can only be constrained observationally with 
large numbers of high-precision radial velocity measurements.

Among the newly discovered ultrafaint dwarf galaxies, Segue 1 has received much 
attention because its proximity and apparently high mass--to--light ratio make it 
an ideal target for indirect dark-matter-detection experiments 
\citep{geha2009,martinez2009,scott2010,essig2010}.  However, for the reasons outlined above, the 
inferred intrinsic velocity dispersion may be susceptible to systematic biases \citep{niederste2009}. A 
confident assessment of these biases requires a larger data set, with repeat 
velocity measurements and an in-depth study of membership issues, contamination 
by streams, and the contribution to the dispersion from binary orbital motions.
In this paper, we undertake this task using the spectroscopic sample of stars 
presented in \citet[][hereafter Paper I]{simon2010}, which also contains the main results of our 
work.  In the present companion paper, we describe in detail our methodology and 
the results pertaining to the intrinsic velocity dispersion of Segue~1. We 
emphasize, though, that the methodology is general and can be applied to 
any dispersion-supported system such as dwarf spheroidal satellites and 
globular clusters.

As a motivation for the methods to be discussed in this paper, we highlight two 
crucial issues. The first is related to velocity outlier stars. The 
analysis of small data sets with a few tens to $\sim100$ stars, typical of 
ultrafaint dwarfs, is always susceptible to large changes due to the 
inclusion or exclusion of certain outliers. For example, in the present Segue~1 
sample the exclusion of one star (SDSSJ100704.35+160459.4) with an intermediate 
membership probability reduces the maximum likelihood velocity dispersion by
$\sim30$\%.  A fully Bayesian analysis does not suffer from this drawback, as 
we explicitly show in this paper.

The second issue is related to repeat measurements with variable measurement 
errors.  Among the brightest and best-studied stars in the Segue 1 sample, the 
six red giants and two horizontal branch stars, there are at least three radial 
velocity variables.  Two of these we identify as RR Lyrae variables, but the 
third appears very likely to be a binary
star, and two additional giants show some ($<2\sigma$) 
evidence for velocity changes as well.  Although the number of stars with 
multiple high-quality velocity measurements is small, the observed variability
of the red giant branch (RGB) stars may be larger than what would be expected if the binary 
population were similar to that of the Milky Way field.  This raises the 
concern that Segue 1 could have a high fraction of binary stars with periods 
short enough ($\lesssim$ 10 years) to inflate the observed velocity dispersion 
significantly.

A recent study by \cite{minor2010} showed that for dwarf galaxies with 
multi-epoch samples of $\approx$ 100 or more stars, 
the binary contribution is unlikely to inflate the
inferred velocity dispersion
by more than 30\%. They also provide a 
method to correct the velocity dispersion for binaries using multi-epoch data.  
In the case of Segue~1, however, the confirmed member sample is 71 stars 
(complete down to $r=21.7$; Paper I), roughly half of which have 
multi-epoch measurements at the present time.  
Two of these members are RR Lyrae variables,
which undergo large velocity variations and therefore should not be
used in the dispersion calculation, leaving 69 members for our
purposes.
Further, the vast majority of 
the sample is made up of main-sequence stars for which the measurement 
errors are quite
large, averaging $\approx$~5.5~km~s$^{-1}$, making the inferred dispersion less 
robust.  The large errors also compound the difficulty of constraining the 
nature of the binary population, since the non-Gaussian tail in the 
line-of-sight velocity 
distribution produced by short-period binaries can be effectively hidden by 
large measurement errors.  Owing to the small multi-epoch sample and the large 
and variable measurement errors, the binary correction given in 
\cite{minor2010} cannot be straightforwardly applied to the Segue~1 data set.

We therefore extend the work of \cite{minor2010} and consider the full 
likelihood for multi-epoch velocity measurements. Along with this extension, we 
introduce a new method to constrain the velocity dispersion of ultrafaint 
dwarf spheroidal galaxies by a comprehensive Bayesian analysis. We apply this 
method to an essentially complete spectroscopic sample of stars within a radius of about 70 
pc from the center of Segue 1 as described in detail in Paper I, and infer the 
intrinsic dispersion of Segue 1. We find with high  confidence that Segue 1 has
a large intrinsic dispersion ($\sim$4 km~s$^{-1}$) as originally estimated by 
\cite{geha2009}, despite evidence of its binary population having shorter 
periods than those observed in the solar neighborhood. 

In our method, we model the multi-epoch likelihood of foreground Milky Way 
stars and both binary and non-binary stars within Segue 1. This likelihood uses 
velocity, metallicity, position, and magnitude information to help determine 
membership and binary properties. In contrast to previous methods,  our 
calculation does not require determining membership probabilities \emph{a 
priori} -- they are implicit in the calculation. It has the additional benefit 
that constraints on the galaxy's binary population can be obtained 
simultaneously with the velocity dispersion. Furthermore, by adding  more 
parameters, our Bayesian analysis can be easily extended to constrain  other 
quantities of interest, e.g., the mass contained within a given radius or  
the galaxy's dark matter annihilation signal. 

Our method can also be used to investigate the presence of additional 
populations (e.g., an overlapping stream, or the presence of distinct stellar 
populations in a dSph). Our preliminary analysis along these lines has not 
revealed any evidence for multiple populations in Segue 1, although the data 
also cannot rule out that possibility.  In addition, allowing for the possibility 
that the stellar velocities are drawn from a dwarf spheroidal 
plus a separate stream-like population
has no significant effect on the inferred intrinsic dispersion.

In Section \ref{sec:membership_method} we will derive a likelihood for both 
member stars and foreground Milky Way stars. In Section \ref{sec:full_method} we 
will derive a multi-epoch likelihood for binary stars and show how this can be 
generated by a Monte Carlo simulation. In Section \ref{sec:priors} we discuss 
our priors on the binary population and how they affect the derived binary 
constraints of Segue 1. The inferred velocity dispersion using this method is 
given in Section \ref{sec:dispersion_results}, and the constraints on Segue 1's 
binary population are discussed in Section \ref{sec:mean_period_results}. In 
Section \ref{sec:stream} we discuss the possibility of contamination by the 
Sagittarius tidal stream, and conclude in Section \ref{sec:conclusion}.

\section{Bayesian Method: Incorporating Membership}\label{sec:membership_method}
\begin{figure*}
\rotatebox{270}{\includegraphics[height=0.482\hsize]{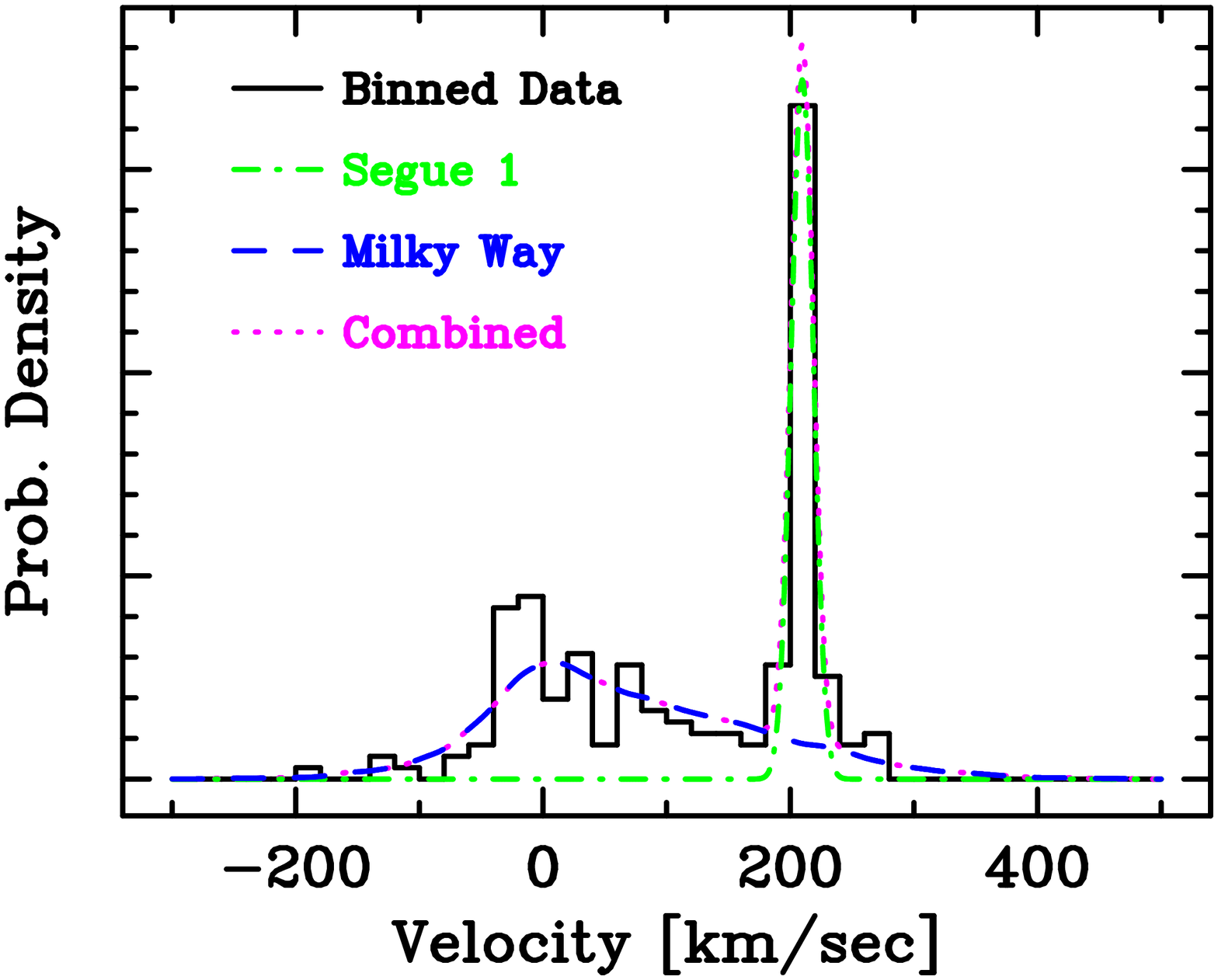}}
\rotatebox{270}{\includegraphics[height=0.482\hsize]{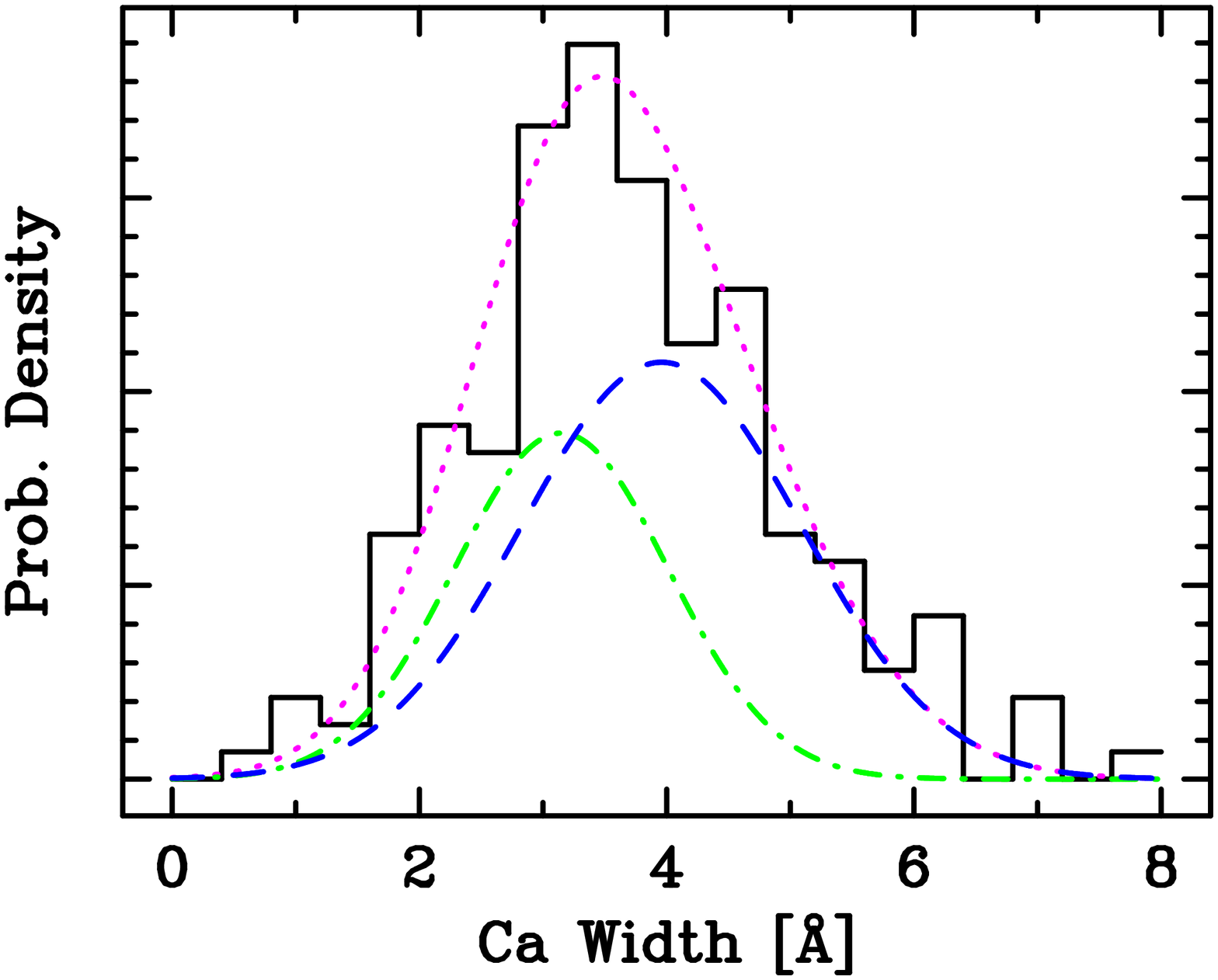}}
\caption{\label{fig:binneddata}\footnotesize 
Distributions of the 
complete Segue 1 data set in line-of-sight velocity (left) and reduced
calcium triplet equivalent width, a proxy for metallicity (right).
We infer the velocity dispersion of Segue 1 by fitting the combined probability 
distribution function (dotted magenta line), composed of both the Milky Way 
(dashed blue line) and Segue 1 (dash-dotted green line) distributions, to the 
complete data set (solid black line); this eliminates the requirement to 
determine the membership of each star a priori. The above graphs illustrate a high 
likelihood parameter set that describes the data well. These parameters are 
marginalized over to obtain probability density functions of relevant model 
parameters (e.g., dispersion and half-light radius).}
\end{figure*}
Membership determination is crucial in estimating Segue 1's 
dynamical properties because the inclusion or exclusion of stars from the proposed 
Segue 1 sample may drastically affect the derived constraints. The most striking 
example is the star SDSSJ100704.35+160459.4, which is a 6$\sigma$ velocity 
outlier but has a relatively high probability of membership due to its close 
proximity to the projected center of Segue 1.  If this star is assumed to be a 
member, the inferred maximum-likelihood velocity dispersion increases from 
$\sigma \approx 4.0$ km~s$^{-1}$ to $\sigma \approx 5.5$ km~s$^{-1}$ (Paper I).  

The most sophisticated and statistically correct method of membership determination 
described so far is the expectation maximization algorithm of \cite{walker11-09}, with the 
primary aim of determining membership probabilities for stars.  We extend this 
method in two essential ways---to allow for parameter space exploration and
parameter estimation. As with the Walker et al. method (illustrated in 
Figure \ref{fig:binneddata}), we do so by modeling  {\em both} the Milky Way
and Segue 1 and simultaneously constraining the model parameters
using the complete data set. Since the membership probabilities are naturally 
incorporated into the analysis, this approach obviates the need to directly evaluate the 
membership of individual stars.

Let us suppose that in the (largely) magnitude-limited Segue 1 sample, a fraction $F$ of 
the stars are members.  To eliminate obvious nonmember stars, a color--magnitude 
cut around the best-fit isochrone is made; spectroscopic measurements are then 
obtained for the remaining stars. For simplicity, we start with the assumption 
that each star has a single velocity measurement $v$ and reduced equivalent
width (EW) of Ca{\textsc{ ii}} triplet absorption lines $w$ (metallicity indicator, see Paper I). If
multiple measurements are made, the velocities and  
metallicities in the following  formulas can be replaced by their average 
values over multiple epochs, suitably  weighted by the measurement errors
as described further in Section \ref{sec:full_method} (see
Equations~(\ref{eq:weighted_average}) and
(\ref{eq:equivalent_error})).  For each star we 
define $R$ to be its projected radius from the center of Segue 1.
The center obtained from SDSS photometry~\citep{martin2008}
is offset by about 32$^{\prime\prime}$ from the mean stellar position of
our spectroscopic sample within 10$^{\prime}$.  However,  
we ran our full analysis by changing the center from the SDSS
photometry to the mean stellar position of our sample and found that
this had little effect on the posterior of the intrinsic
velocity dispersion. Therefore, for the rest of the analysis, we fix
the center to the SDSS photometry value.
Assuming there are only two stellar populations, the 
Milky Way (MW) and Segue 1 (gal) galaxies, the joint likelihood for a single data point 
$\mathscr{D}_i = \{v, e, w, e_w, R\}_i$ is 
\begin{equation}
 \mathcal{L}(\mathscr{D}_i \vert \mathscr{M}) = F \mathcal{L}_{\mathrm{gal}}(\mathscr{D}_i \vert \mathscr{M}_{\mathrm{gal}})
  +(1-F) \mathcal{L}_{\mathrm{MW}}(\mathscr{D}_i \vert \mathscr{M}_{\mathrm{MW}}).
\end{equation}
Here, $\mathcal{L}_{\mathrm{gal}}$ and $\mathcal{L}_{\mathrm{MW}}$ are the individual probability 
distributions of Segue 1 and the Milky Way parameterized by the sets 
$\mathscr{M}_{\mathrm{gal}, \mathrm{MW}}$.  All sources of measurement error in $v$ and
$w$ are included in $\epsilon$ and $\epsilon_w$, respectively, and we
model the measurements as being drawn from a Gaussian distribution
with these errors. The metallicity distributions of the member and 
nonmember stars are each modeled by Gaussians with mean metallicities $\bar 
w_{\mathrm{gal}}$, $\bar w_{\mathrm{MW}}$ and widths $\sigma_{w, \mathrm{gal}}$, $\sigma_{w,\mathrm{MW}}$ 
respectively. 
We assume that metallicity has no spatial or
  velocity dependence because no metallicity gradients have been
  detected in any of the ultrafaint dwarfs. The likelihood is
assumed to be separable in velocity, position,  
and metallicity, so that each individual probability distribution can now be 
written as
\begin{equation} \label{eq:lgalmw}
\mathcal{L}_{\mathrm{gal}, \mathrm{MW}}(v, w, R) = \mathcal{L}_{\mathrm{gal}, \mathrm{MW}}(w) \mathcal{L}_{\mathrm{gal}, \mathrm{MW}}(v \vert R) \mathcal{L}_{\mathrm{gal}, \mathrm{MW}}(R)\textrm{,}
\end{equation}
where
\begin{equation} \label{eq:lgalw}
\mathcal{L}_{\mathrm{gal}, \mathrm{MW}}(w) = \frac{1}{\sqrt{2\pi \sigma_{w, \mathrm{gal}, \mathrm{MW}}^2}} \exp \left[-\frac{(w - \bar w_{\mathrm{gal}, \mathrm{MW}})^2}{2 \sigma_{w, \mathrm{gal}, \mathrm{MW}}^2}\right].
\end{equation}
We have momentarily dropped the model parameter notation $\mathscr{M}$ for clarity.  The 
last factor in Equation (\ref{eq:lgalmw}) has a simple physical interpretation:  the 
spatial probability distribution is the projected number (surface) density of stars normalized to 
unity.  Note however that this surface density is the density of \emph{observed} 
stars, which may heavily be influenced by selection biases.  Thus, we write the 
observed spatial probability density as $\mathcal{L}(R) = n(R)\mathcal{S}(x, 
y)/N$, where $n(R)$ is the actual surface density of the member stars, $N$ is 
the total number of stars in the sample, and $\mathcal{S}(x,y)$ 
represents any bias introduced by observational selection.
In the classical dSphs, which contain hundreds to 
thousands of bright member stars, the selection function may be difficult to 
quantify, but in the much sparser ultrafaints it is often more 
straightforward to model the spectroscopic selection \citep{Willman2010, 
simon2010}. To avoid spatial selection biases, we use the conditional 
likelihood $\mathcal{L}(v, w \vert R) = \mathcal{L}(v, w, R)/\mathcal{L}(R)$.  
From the previous discussion, we have
\begin{eqnarray} \label{eq:condlike}
 \mathcal{L}(v, w \vert R) & = & f(R)\mathcal{L}_{\mathrm{gal}}(w)\mathcal{L}_{\mathrm{gal}}(v \vert R) \nonumber \\
 & & + \left(1 - f(R)\right) \mathcal{L}_{\mathrm{MW}}(w)\mathcal{L}_{\mathrm{MW}}(v \vert R)
\label{eq:likelihood_vwr}
\end{eqnarray}
where $f(R)$ is the fraction of stars that are dwarf galaxy members at the position $R$:
\begin{equation} \label{eq:fr}
f(R) = \frac{n_{\mathrm{gal}}(R)}{n_{\mathrm{gal}}(R) + n_{\mathrm{MW}}(R)}.
\end{equation}
In principle, the selection bias affects the Milky Way and dSph distributions 
equally, so that by Equation~(\ref{eq:fr}) 
the membership fraction $f(R)$ should be insensitive to these
selection biases.  Put another way, the spatial selection bias only
affects the total number of stars  selected and not the fraction of
those stars that are  members. 

The Segue 1 data set is fairly unique in that, within the
given color, magnitude, and spatial cuts, the sample is essentially
complete up to its magnitude limit of $r = 21.7$, although it does
also extend to somewhat fainter  magnitudes and larger radii (Paper I). 
Thus, spatial selection biases are not expected to be significant and
we may also use the full likelihood, $\mathcal{L}(v, w, R)$,
incorporating the spatial dependence directly.  The conditional
likelihood (Equation~(\ref{eq:condlike})) is better suited for situations
where the spectroscopic data set is not complete, which is more
typical. We find that the inferred velocity dispersion 
of Segue 1 is insensitive to whether we use the full likelihood
or the conditional likelihood.  However, as shown below, the full
likelihood does provide a tighter 
constraint on the stellar distribution itself. The results obtained
from these two methods are compared in Table \ref{tab:params}. We also
list the priors used for each parameter in that table.  
In this paper we use the conditional likelihood (Equation~(\ref{eq:condlike}))
by default unless the positional information becomes important. This
is the case for the half-light radius and the inferred dark matter
density with the half light radius, for which we quote results
obtained from the full likelihood (Equation~(\ref{eq:lgalmw})).

The projected number density of the dSph stars is modeled by a
modified Plummer profile of the form
\begin{equation} \label{eq:modplum}
 n_{\mathrm{gal}}(R) \propto \left(1 + (R/R_s)^2\right)^{-(\alpha-1)/2},
\end{equation}
where $\alpha=5$ is the standard Plummer profile integrated along the line
of sight. 
Using the conditional likelihood, the data are not able to
constrain the outer slope ($\alpha$).  
But the full likelihood analysis does provide a modest constraint, 
$\alpha=4.1^{+2.0}_{-0.8}$, which is consistent with a Plummer profile
(see Table \ref{tab:params}). The number density
of Milky Way stars is assumed to be spatially constant over the field of view,
which should be a reasonable approximation for compact systems such as
Segue 1.  The normalization of the Milky Way likelihood in $R$,  
which we call $n_{\mathrm{MW}, 0}$, is thus determined solely by the cutoff 
radius, which we take to be that of the star farthest from the center of the 
galaxy.  For determining membership, however, only the relative normalization 
between the dSph and Milky Way number densities is important in
Equation (\ref{eq:fr}); this is given by  
\begin{equation} \label{eq:R}
N_{\mathrm{gal}} = \frac{n_{\mathrm{gal}}(0)}{n_{\mathrm{MW}, 0}}\,. 
\end{equation}
We therefore include $N_{\mathrm{gal}}$ as a model parameter.

Neglecting binaries, the velocity distribution of Segue 1 is assumed to be 
Gaussian with dispersion $\sigma$ and mean velocity $\mu$.  Although in 
principle any velocity distribution can be used, there is currently no evidence 
for large deviations from Gaussianity in dSph velocity distributions \citep[e.g.][]{walker2006}.
In Section \ref{sec:full_method}, we discuss how this velocity distribution is 
modified by the presence of binary stars. For the velocity likelihood of Milky 
Way stars, we use the Besancon model (\citealt{robin2003}) together with the 
appropriate color-magnitude cuts.  However, to allow for uncertainties in the 
Besancon model, we allow the velocity distribution to be shifted by a small 
amount $\delta$ and stretched by a factor $S$, both are shown to be 
well determined by the data. 
We also explored the potential
  effects of assuming other foreground models---a ``noisy'' Besancon
  model  and a Gaussian fit whose peak is offset by about 
  50 km~s$^{-1}$---but found no significant effect on the inferred posterior
  for the intrinsic dispersion. We therefore do not discuss these alternate
foreground models further. 

Our resulting set of 14 model parameters is
\begin{equation}
\mathscr{M} = \{N_{\mathrm{gal}},\sigma,\mu,\bar w, \sigma_w, \bar w_{\mathrm{MW}}, \sigma_{w,\mathrm{MW}}, 
R_s, \delta, S, \alpha \}.
\label{eq:lots_of_parameters}
\end{equation}

The probability density of the model parameters $\mathscr{M}$
given the data sets $\mathscr{W} = \{w_i\}$, $\mathscr{V} = \{v_i\}$, and 
$\mathscr{R} = \{R_i\}$ can now be written as
\begin{equation}
\mathcal{P}({\mathscr{M} \vert \mathscr{W}, \mathscr{V}. \mathscr{R}}) \propto
\mathcal{L}( \mathscr{W}, \mathscr{V} | \mathscr{R}, \mathscr{M}) \mathcal{P}(\mathscr{M})\textrm{,}
\end{equation}
where $\mathcal{L}( \mathscr{W}, \mathscr{V} | \mathscr{R}, \mathscr{M}) 
= \prod_i \mathcal{L}( w_i, v_i | R_i, \mathscr{M})$ is the likelihood function
for the complete data set and $\mathcal{P}(\mathscr{M})$ is the prior on the 
model parameters. We choose uniform priors in the above parameters with the 
exception of the metallicity distribution widths, $\sigma_w$ and 
$\sigma_{w,\mathrm{MW}}$, for which we choose the usual non-informative priors that are 
uniform in log-space. To conservatively bias our member probabilities (and 
consequently the dispersion) low, we choose the $N_{\mathrm{gal}}$ and $R_s$ priors also to be 
uniform in log-space; however, we found the form of the priors in these 
parameters to have little effect on the inferred dispersion.  The prior on 
velocity dispersion was chosen to be uniform since this is the parameter of 
interest.

After estimating the model parameters $\mathscr{M}$, we can derive membership 
probabilities for each individual star. The formula for the probability of 
membership for the $i$th star is
\begin{equation}
p_i = \frac{f(R_i) \mathcal{L}_{\mathrm{gal}}(w_i, v_i \vert R_i)}
{f(R_i) \mathcal{L}_{\mathrm{gal}}(w_i, v_i \vert R_i) 
+ \left(1 - f(R_i)\right) \mathcal{L}_{\mathrm{MW}}(w_i, v_i \vert R_i)}.
\end{equation}
Because we derive a probability distribution in the model parameters 
$\mathscr{M}$, the probability distribution for $p_i$ can be obtained
using our method.  Here, we will quote the average membership
probability $\left<p_i\right>$. 
\begin{sidewaystable*}\label{tab:params}
\caption{\label{tab:limits}Summary of Model Parameters and the Priors Assumed}
\normalsize
\begin{center}
\begin{tabular}{@{}ccccl}
\noalign{\hrule height 1pt}
\vspace{1pt}
Velocity Distribution & & Derived Value & Derived Value \\
Parameters & Priors Assumed\tablenotemark{\dag} & Conditional Likelihood\tablenotemark{1} &
Full Likelihood\tablenotemark{2} & Description\\ 
\hline
\vspace{1pt}
$\sigma$ & $0$ km~s$^{-1}$ $ < \sigma < 10$ km~s$^{-1}$ & $3.7^{+1.4}_{-1.1}$ km~s$^{-1}$ &
$3.5^{+1.8}_{-1.0}$ km~s$^{-1}$& Intrinsic velocity dispersion of Segue 1\\ 
$\mu$ & $200$ km~s$^{-1}$ $< \mu < 220$ km~s$^{-1}$  & $209^{+1}_{-1}$ km~s$^{-1}$ &
$209^{+1}_{-1}$ km~s$^{-1}$ & Systemic velocity of Segue 1\\ 
${\bar w}_{\mathrm{gal}}$ & $2$ $ \mathrm{\AA}$ $< {\bar w} < 6 $ $\mathrm{\AA}$ &
$3.1^{+0.2}_{-0.1}$ $\mathrm{\AA}$ & $3.2^{+0.1}_{-0.2}$ $\mathrm{\AA}$& Segue 1 average
reduced Ca EW  (Equation (\ref{eq:lgalw}))\\ 
$\sigma_{w, \mathrm{gal}}$ & $-2 < \log_{10}(\sigma_w[\mathrm{\AA}]) < 1$ &
$0.05^{+0.07}_{-0.06}$ & $0.03^{+0.06}_{-0.07}$ & Segue 1 reduced Ca EW dispersion (Equation (\ref{eq:lgalw}))\\
$\bar w_{\mathrm{MW}}$ & $2$ $ \mathrm{\AA}$ $< \bar w < 6 $ $\mathrm{\AA}$ & $4.0^{+0.1}_{-0.1}$
$\mathrm{\AA}$ & $4.0^{+0.1}_{-0.1}$ $\mathrm{\AA}$& MW average reduced Ca EW width (Equation (\ref{eq:lgalw}))\\
$\sigma_{w, \mathrm{MW}}$ & $-2 < \log_{10}(\sigma_w[\mathrm{\AA}]) < 1$ &
$0.06^{+0.03}_{-0.04}$ & $0.05^{+0.04}_{-0.04}$ & MW reduced Ca EW
dispersion (Equation (\ref{eq:lgalw}))\\ 
$\delta$ & $-70$ km~s$^{-1}$ $< \delta < 10$ km~s$^{-1}$ & $-19^{+7}_{-8}$ km~s$^{-1}$ &
$-20^{+6}_{-9}$ km~s$^{-1}$ & Shift in the MW velocity distribution\\ 
$S$ & $-2 < \log_{10}(S) < 1$ & $0.03^{+0.05}_{-0.05}$ &
$0.01^{+0.05}_{-0.05}$ & Scale in the MW velocity distribution
\vspace{1pt}\\
\noalign{\hrule height 1pt}
\\
\noalign{\hrule height 1pt}
\vspace{1pt}
Stellar Profile & & Derived Value & Derived Value \\
Parameters & Priors Assumed\tablenotemark{\dag} & Conditional Likelihood\tablenotemark{1} &
Full Likelihood\tablenotemark{2} & Description\\ 
\hline
\vspace{1pt}
$R_s$ & $1 < \log_{10}(R_s[$pc$]) < 2$ & $1.8^{+0.1}_{-0.4}$ & $1.4^{+0.2}_{-0.2}$ & Scale radius (Equation (\ref{eq:modplum}))\\
$\alpha$ & $3 < \alpha < 10$ & \textasteriskcentered & $4.1^{+2.0}_{-0.8}$ & Outer log slope (Equation (\ref{eq:modplum}))\\
$N_{\mathrm{gal}}$ & $-1 < \log_{10}(N_{\mathrm{gal}}) < 3$  & $0.5^{+0.3}_{-0.2}$ &
$1.0^{+0.2}_{-0.2}$ & Segue 1 central density / MW density
(Equation (\ref{eq:R})) 
\vspace{1pt}\\
\noalign{\hrule height 1pt}
\\
\noalign{\hrule height 1pt}
\vspace{1pt}
 & & Derived Value & Derived Value \\
Binary Parameters & Priors Assumed\tablenotemark{\dag} & Conditional
Likelihood\tablenotemark{1} & Full Likelihood\tablenotemark{2} & Description\\
\hline
\vspace{1pt}
$B$ & $0 < B < 1$ & \textasteriskcentered & \textasteriskcentered & Binary fraction\\
$\sigma_{\log_{10}(P)}$ & $0.5 < \sigma_{\log_{10}(P)} < 2.3$ &
\textasteriskcentered & \textasteriskcentered & Dispersion of the orbital period distribution\\ 
$\mu_{\log_{10}(P)}$ & MW composite prior (see the text) &
$1.2^{+1.2}_{-1.2}$ & $0.8^{+1.0}_{-1.0}$ & Mean of the orbital period distribution
\vspace{1pt}\\
\noalign{\hrule height 1pt}
\end{tabular}
\end{center}
{\noindent \ $^{\textrm{\textasteriskcentered}}$\footnotesize Value not constrained.}\\
{\indent \ $^{\textrm{\dag}}$\footnotesize Unless otherwise stated, the prior is assumed to be flat within the given range.}
\tablenotetext{1}{Using the conditional likelihood
  $\mathcal{L}(\mathscr{V}, \mathscr{W} \vert \mathscr{R})$ given by 
  Equation (\ref{eq:condlike})}
\tablenotetext{2}{Using the likelihood
  $\mathcal{L}(\mathscr{V}, \mathscr{W}, \mathscr{R})$ given by 
  Equation (\ref{eq:lgalmw}) where we include the spatial information
  directly. \\Note that the results from 1 and 2 are 
  quantitatively similar. The differences arise due to the fact that
  the half-light radius (which is determined by both $R_s$ and
  $\alpha$) is better constrained by using the full likelihood. Except
  for the constraints on the half-light radius and the dark matter
  density within the half-light radius, our final results are
  based on the (more conservative) conditional likelihood. }  
\end{sidewaystable*}

\section{Bayesian method: correcting for binaries}\label{sec:full_method}

Apart from contamination by nonmember stars, the observed velocity dispersion of Segue 
1 may also be inflated by binary orbital motion. One method of correcting the 
dispersion for binary motion is given in \cite{minor2010}. This method requires 
measuring the threshold fraction of the sample, defined as the fraction of 
stars with observed change in velocity greater than a certain threshold after a 
time interval (typically ~1 year). Provided that velocity outlier stars are 
discarded when determining the dispersion (e.g., by a 3$\sigma$ clip), the 
threshold fraction $F$ is tightly correlated with the dispersion introduced by 
binaries. This relation can be used to correct the dispersion for binaries.  
Although the threshold fraction is defined in terms of two epochs, it can be 
better determined using more than two epochs with a likelihood approach.  This 
approach also has the advantage that it uses only velocity changes to 
characterize the binary population, and hence is less affected by contamination 
by nonmember stars than if the velocities were used directly.

Unfortunately, this method is not ideal for the present data
set of ultrafaint galaxies like Segue 1 for several reasons. First,
the majority of the sample consists of  faint main-sequence stars (and
not red giants) for which the measurement errors are considerable (of
the  same order as  the dispersion itself).  Given this and the
present sample size for Segue 1 (65 stars with multi-epoch
measurements, roughly half of which are members), the threshold
fraction is not well determined.  
Second, the relation between threshold fraction and 
dispersion is a result of the degeneracy of binary fraction with other 
properties characterizing the binary population (e.g., mean period). However, 
this degeneracy is weaker for main-sequence stars than for red giants, so that 
the uncertainty in the binary correction of \cite{minor2010}  becomes wider by 
a factor of ~two, though it is mainly at the small dispersion end.  
Third, this method only corrects the dispersion by an amount that is the same 
for each star, whereas individual stars with large observed velocity changes 
should in principle receive a larger correction.

While the \citet{minor2010} method can still be applied, we adopt a
more ambitious approach: modeling the multi-epoch  likelihood of
binary stars and incorporating it into a comprehensive Bayesian
analysis. In this approach, we include as model parameters the binary
fraction  $B$, mean period $\mu_{\log P}$, and width of the period
distribution $\sigma_{\log P}$. Since the individual velocities are
used, in order to  distinguish between binaries and nonmember stars
we will also need to model  
the likelihood of nonmember stars as in the previous section. In principle 
this is the best possible method for determining the intrinsic dispersion of a 
dwarf galaxy or cluster, since it uses all the available information to 
constrain properties of the binary, member, and nonmember populations in a 
consistent way.

\subsection{Multi-epoch likelihood}\label{subsec:likelihood}

In order to correct the velocity dispersion of dwarf spheroidal galaxies for 
binaries, we must extend the Bayesian method developed in Section 2 
to include the effect of binary stars. First we neglect the Milky Way component 
and focus on the dwarf galaxy likelihood, for which the dynamical parameters 
are the velocity dispersion ($\sigma$) and systemic velocity ($\mu$). We take as a model 
parameter the fraction ($B$) of the stars in binary systems, and we further 
model the binary population by a set of parameters $\mathscr{P}$ that 
characterize the distributions of binary properties. In general, these binary 
properties may include the periods, mass ratios and orbital eccentricities.  
The distributions of these properties and our choice of model parameters will 
be discussed in detail in Section \ref{sec:priors}.

Suppose a star of absolute magnitude $M$ has a set of $n$ velocity measurements 
$\{v_i\}=\{v_1,\dots,v_n\}$ and errors $\{e_i\}$ taken at the 
corresponding dates $\{t_i\}$. For readability, when denoting probability 
distributions we will suppress the brackets denoting sets of measurements 
(e.g., $P(\{v_i\}) \rightarrow P(v_i)$). For reasons that will become clear 
later, we will write the likelihood of each star in terms of a \emph{joint} 
probability distribution in the measured velocities $v_i$ and $v_{\mathrm{cm}}$, the 
velocity of the star system's center of mass (which is unknown), and then 
integrate over $v_{\mathrm{cm}}$.  The likelihood can be written as
\begin{eqnarray}
\lefteqn{
\mathcal{L}(v_i|e_i,t_i,M; \sigma,\mu,B,\mathscr{P}) } \nonumber \\
&=& \int_{-\infty}^{\infty} P(v_i,v_{\mathrm{cm}}|e_i,t_i,M; \sigma, \mu, B, 
\mathscr{P})dv_{\mathrm{cm}} \nonumber \\
&=& \int_{-\infty}^{\infty} P(v_i|v_{\mathrm{cm}},e_i,t_i,M;B,\mathscr{P}) 
P(v_{\mathrm{cm}}|\sigma,\mu)dv_{\mathrm{cm}}.
\label{eq:joint_like}
\end{eqnarray}
The second factor in the integrand is the probability distribution of the 
center-of-mass velocity of the stars, which we take to be Gaussian:
\begin{equation}
P(v_{\mathrm{cm}}|\sigma,\mu) = \frac{e^{-(v_{\mathrm{cm}}-\mu)^2/2\sigma^2}}{\sqrt{2\pi\sigma^2}}
\label{eq:cm_like}
\end{equation}
The first factor in the integrand of Equation~(\ref{eq:joint_like}) is the probability 
of drawing a set of velocity measurements $\{v_i\}$ given a star with
center-of-mass velocity $v_{\mathrm{cm}}$.  This probability distribution is determined 
by two factors, binarity and measurement error.  It can be written as follows:
\begin{eqnarray}
\lefteqn{P(v_i|v_{\mathrm{cm}},e_i,t_i,M;B,\mathscr{P})} \nonumber \\
& = & (1-B)\prod_{i=1}^n \frac{e^{-(v_i-v_{\mathrm{cm}})^2/2e_i^2}}{\sqrt{2\pi e_i^2}} + B P_b(v_i|v_{\mathrm{cm}},e_i,t_i,M;\mathscr{P}) \nonumber \\
& = & (1-B)\mathcal{N}(v_i,e_i) \frac{e^{-(v_{\mathrm{cm}} - \langle v 
\rangle)^2/2e_m^2}}{\sqrt{2\pi e_m^2}} \nonumber\\
& & ~~~~~~~~~ + ~ B P_b'(v_i-v_{\mathrm{cm}}|e_i,t_i,M;\mathscr{P})
\label{eq:v_given_vcm_like}
\end{eqnarray}
where $P_b'(v_i-v_{\mathrm{cm}}|e_i,t_i,M;\mathscr{P})$ is the likelihood in the center-of-mass 
frame of the binary system, with the velocity in the center-of-mass frame given by $v_i' 
= v_i - v_{\mathrm{cm}}$.  In the first term, $\langle v\rangle$ and $e_m$ are the 
weighted average velocity and equivalent measurement error,
\footnote{The data $e_i$ in Equations~(\ref{eq:v}) and~(\ref{eq:e}) include all the
  sources of error identified in \citet{simon2007}. These errors could have a 
  systematic component (as suggested by \citealt{simon2007}) that does not 
  average out statistically as we have assumed here. A greater number of repeat
  independent velocity measurements would be required to test for this
  scenario. For consistency, the Calcium triplet EWs are
  averaged in exactly the same manner as the velocity
  measurements.}     
\begin{equation} \label{eq:v}
\langle v \rangle = e_m^2 \sum_{i=1}^n \frac{v_i}{e_i^2},
\label{eq:weighted_average}
\end{equation}
\begin{equation} \label{eq:e}
e_m^2 = \left(\sum_{i=1}^{n}\frac{1}{e_i^2}\right)^{-1},
\label{eq:equivalent_error}
\end{equation}
while the normalizing factor $\mathcal{N}$ is given by
\begin{eqnarray}
\lefteqn{
\mathcal{N}(v_i,e_i) ~ ~ = ~ ~ \frac{\sqrt{2\pi e_m^2}}{\prod_{i=1}^n \sqrt{2\pi e_i^2}} } \nonumber\\
& \times & \exp\left\{-\frac{1}{4}\sum_{i,j=1}^n\frac{(v_i-v_j)^2}{e_i^2 + 
e_j^2 + e_i^2 e_j^2\left(\sum_{k\neq 
i,j}\frac{1}{e_k^2}\right) } \right\}.
\label{eq:n_factor}
\end{eqnarray}
The last term in the denominator of the exponent is implicitly zero when $n = 
2$.

Multiplying Equation~(\ref{eq:v_given_vcm_like}) by Equation~(\ref{eq:cm_like}) and 
integrating in accordance with Equation~(\ref{eq:joint_like}), we find:
\begin{eqnarray}
\lefteqn{\mathcal{L}(v_i|e_i,t_i,M;\sigma,\mu,B,\mathscr{P})} \nonumber \\
& \propto & ~ (1-B)\frac{e^{-\frac{(\langle v\rangle - \mu)^2}{2(\sigma^2 + e_m^2)}}}{\sqrt{2\pi(\sigma^2 + e_m^2)}} + BJ(\sigma,\mu,\mathscr{P}),
\label{eq:full_binary_likelihood}
\end{eqnarray}
where we have left off the normalizing $\mathcal{N}$ factor, and
\begin{equation}
J(\sigma,\mu,\mathscr{P}) = \int_{-\infty}^{\infty} \mathcal{R}(v_{\mathrm{cm}}) \frac{e^{-\frac{(v_{\mathrm{cm}} - \mu)^2}{2\sigma^2}}}{\sqrt{2\pi\sigma^2}} dv_{\mathrm{cm}},
\label{eq:i_integral}
\end{equation}
\begin{equation}
\mathcal{R}(v_{\mathrm{cm}},\mathscr{P}) = \frac{P_b'(v_i-v_{\mathrm{cm}}|e_i,t_i,M;\mathscr{P})}{\mathcal{N}(v_i,e_i)}.
\label{eq:r_function}
\end{equation}
Since the factor $\mathcal{N}$ is independent of all model parameters, it is 
usually ignored in the likelihood when averaging velocities without regard for 
binaries (i.e., it acts only as a normalizing factor). As 
Equation~(\ref{eq:r_function}) shows, however, it is crucial to include here since it 
determines the relative normalization of the binary and non-binary terms. Note 
that if a star exhibits large velocity variations compared to the measurement 
errors, according to Equation~(\ref{eq:n_factor}) the $\mathcal{N}$ factor will be 
quite small. If in addition the velocity variations are observed over some time 
interval consistent with binary behavior, the normalization of 
$\mathcal{R}(v_{\mathrm{cm}})$ will be greatly enhanced, possibly by orders of 
magnitude, because of the $\mathcal{N}$ factor in the denominator of 
Equation~(\ref{eq:r_function}).

For each star that has multi-epoch data, we run a Monte Carlo simulation and 
bin the velocities over a table of $v_{\mathrm{cm}}$ values to find 
$P_b'(v_i-v_{\mathrm{cm}}|e_i,t_i,M)$. The $\mathcal{R}$-function is recorded for 
each star and subsequently integrated to evaluate $J(\sigma,\mu,\mathscr{P})$.

\subsection{Binary population model uncertainties}\label{sec:priors}

To infer the intrinsic velocity dispersion of Segue 1, we must marginalize over 
the parameters characterizing the binary population. It is therefore critical 
to address the question of which binary model parameters to use and how to deal 
with uncertainties in these parameters. Besides the binary fraction, a 
population of binary stars can be described by distributions in three 
parameters: the mass ratio $q$, eccentricity $e$, and orbital period $P$. In 
the absence of a large number of epochs, eccentricities are difficult to 
constrain because very eccentric binaries spend a relatively small amount of 
time near their perihelion where the observed velocities are large. We 
therefore fix the distribution of eccentricities and assume the form given in 
\cite{minor2010}, which is similar to that observed in solar neighborhood field 
binaries.

Along similar lines, velocity measurements at several epochs are usually needed 
to determine the mass ratio of a binary independently of its orbital period.  
Evidence suggests, however, that the period distributions of different binary 
populations can differ drastically, while the distribution of mass ratios may 
have a more nearly universal form. This is certainly true for long-period 
binaries, for which the mass ratio follows the Salpeter initial mass function 
for $q \gtrapprox 0.5$ (assuming the primary mass to lie in a very restricted 
range, as is the case for the observed sample in Segue 1; 
cf. \citealt{duquennoy1991}).  The observed distribution of mass ratios 
for short-period binaries ($P < 1000$ days) is closer to uniform 
(\citealt{goldberg2003}, \citealt{mazeh1992}), and at present it is unclear 
whether this form is universal in primordial binary populations.  
We therefore fix the mass ratio distribution and assume it to follow a form 
similar to that observed in the solar neighborhood, as described in 
\citet{minor2010}, with a uniform distribution for short-period
binaries. Note that we are allowing for the mass ratio and ellipticity
to vary from star to star---it is just the form of the distribution
from which these parameters are derived that is fixed. 
In principle, we could also vary the functional form of $q$ and $e$,
but this is computationally expensive. The main reason is that the 
function ${\cal R}(v_{\mathrm{cm}},{\mathscr P})$ (see Equation~(\ref{eq:r_function}))
will have to be computed on a grid that includes the parameters used
to describe the functional form of $q$ and $e$ distributions. In
addition, given the small data set and the large measurement errors,
these parameters will be highly degenerate with other binary
parameters ($B$, $\mu_{\log P}$, $\sigma_{\log P}$).  

Although binary populations in open clusters have been observed to display a 
narrower distribution of periods than binaries in the field 
(\citealt{brandner1998}; \citealt{scally1999}), they still range over multiple 
decades of period.  For simplicity we assume the period distribution of Segue 1 
to have a log-normal form, in analogy to field binaries 
(\citealt{duquennoy1991}; \citealt{fischer1992}; \citealt{mayor1992}; \citealt{raghavan2010}), while 
the mean period $\mu_{\log P}$ and spread of periods $\sigma_{\log P}$ will be 
allowed to differ from that observed in solar neighborhood field binaries.  We 
therefore have three binary parameters that are allowed to vary: the binary 
fraction $B$, the mean log-period $\mu_{\log P}$, and log-spread of periods 
$\sigma_{\log P}$.

\begin{figure}
\rotatebox{270}{\includegraphics[height=0.964\hsize]{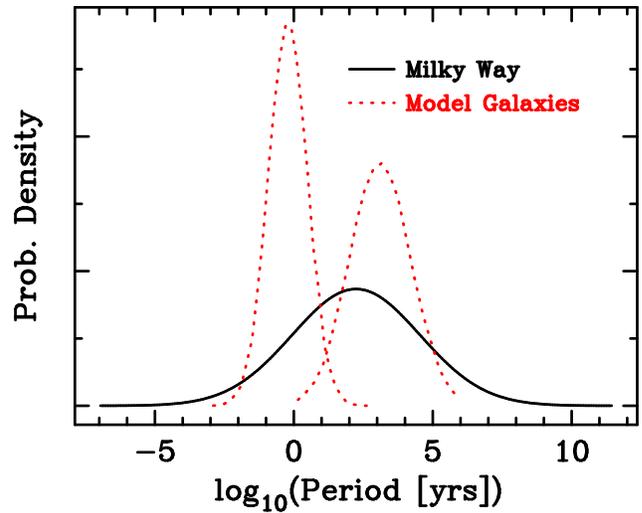}}
\caption{\label{fig:logpdists}\footnotesize Period distributions of simulated 
clusters generated from our priors, compared to the observed period 
distribution of field binaries in the solar neighborhood with solar-mass 
primaries (\citealt{duquennoy1991}). The simulated distributions have widths 
drawn from a flat prior with a range $\sigma_{\log P} \in [0.5,2.3]$, and mean 
periods drawn from a Gaussian prior chosen such that when many cluster period 
distributions are superimposed, they form the \cite{duquennoy1991} period 
distribution of field binaries.}
\end{figure}

Since the binary fraction $B$ may have any value between 0 and 1, we choose a 
uniform prior in $B$ over this interval. Our prior in the period distribution 
parameters $\mu_{\log P}$ and $\sigma_{\log P}$, however, requires more careful 
consideration.  The prevailing viewpoint is that the observed distribution of 
field binary stars in the solar neighborhood is a superposition of populations 
from a wide variety of star-forming environments with different period 
distributions; this is supported by the fact that several clusters seem to 
exhibit period distributions that are more peaked than those observed in the 
field (\citealt{brandner1998}, \citealt{scally1999}). We shall therefore make 
the assumption that the binary populations in dwarf spheroidal galaxies follow 
period distributions that are a subset of the distribution exhibited by solar 
neighborhood field binaries.  Given the fact that the period distribution in 
the solar neighborhood is nearly flat in log-space over the relevant parameter 
space, one option is to use flat priors in $\mu_{\log P}$ and $\sigma_{\log 
P}$. However, the limits of integration are somewhat arbitrary, and may allow 
populations with binary periods shorter or longer than any observed in the 
solar neighborhood. A somewhat better-motivated method is to assume a flat 
prior in $\sigma_{\log P}$ with a certain range $[\sigma_{\log P, \mathrm{min}},2.3]$, 
and then find a prior distribution in the mean period such that when many 
binary populations are drawn from these priors, they superimpose to form the 
\cite{duquennoy1991} period distribution observed in field binaries.  This is 
illustrated in Figure \ref{fig:logpdists} where we plot a few cluster period 
distributions which have been generated from this prior, together with the 
field binary period distribution of \cite{duquennoy1991} observed in the solar 
neighborhood that has parameters $\mu_{\log P} = 2.23$ ($P$ in years) and 
$\sigma_{\log P} = 2.3$.  We assume a Gaussian prior for $\mu_{\log P}$ with a 
mean $\bar \mu_{\log P} = 2.23$, then maximize a likelihood to find the width 
$\sigma_\mu$ of this prior required to reproduce the field binary distribution 
when a large number of populations are superimposed.  If we choose a minimum 
period spread $\sigma_{\log P, \mathrm{min}} = 0.5$, we find that the width of the mean period 
prior satisfying these conditions is $\sigma_\mu = 1.7$. While this
prior already encapsulates a very wide range of period distributions,
we also investigate more extreme priors and show that our inferred
velocity distribution is not significantly affected by our priors. 

\begin{figure*}[h!]
	\subfigure[0.4 km~s$^{-1}$ intrinsic dispersion, 10 year mean period]
	{
		\rotatebox{-90}{\includegraphics[height=0.482\hsize]{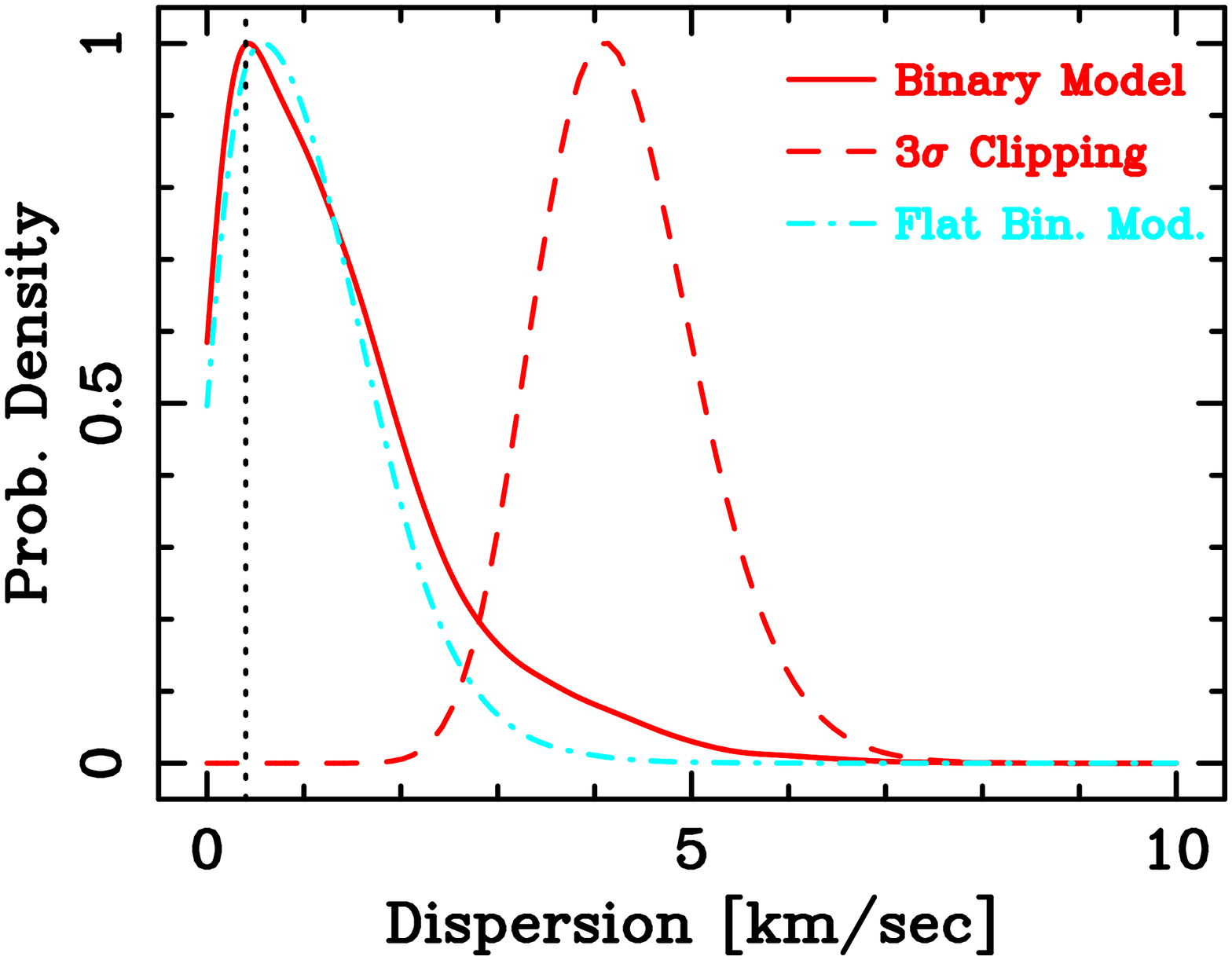}}
		\label{fig:sigtest04}
	}
	\subfigure[0.4 km~s$^{-1}$ intrinsic dispersion, 10 year mean period]
	{
		\rotatebox{-90}{\includegraphics[height=0.482\hsize]{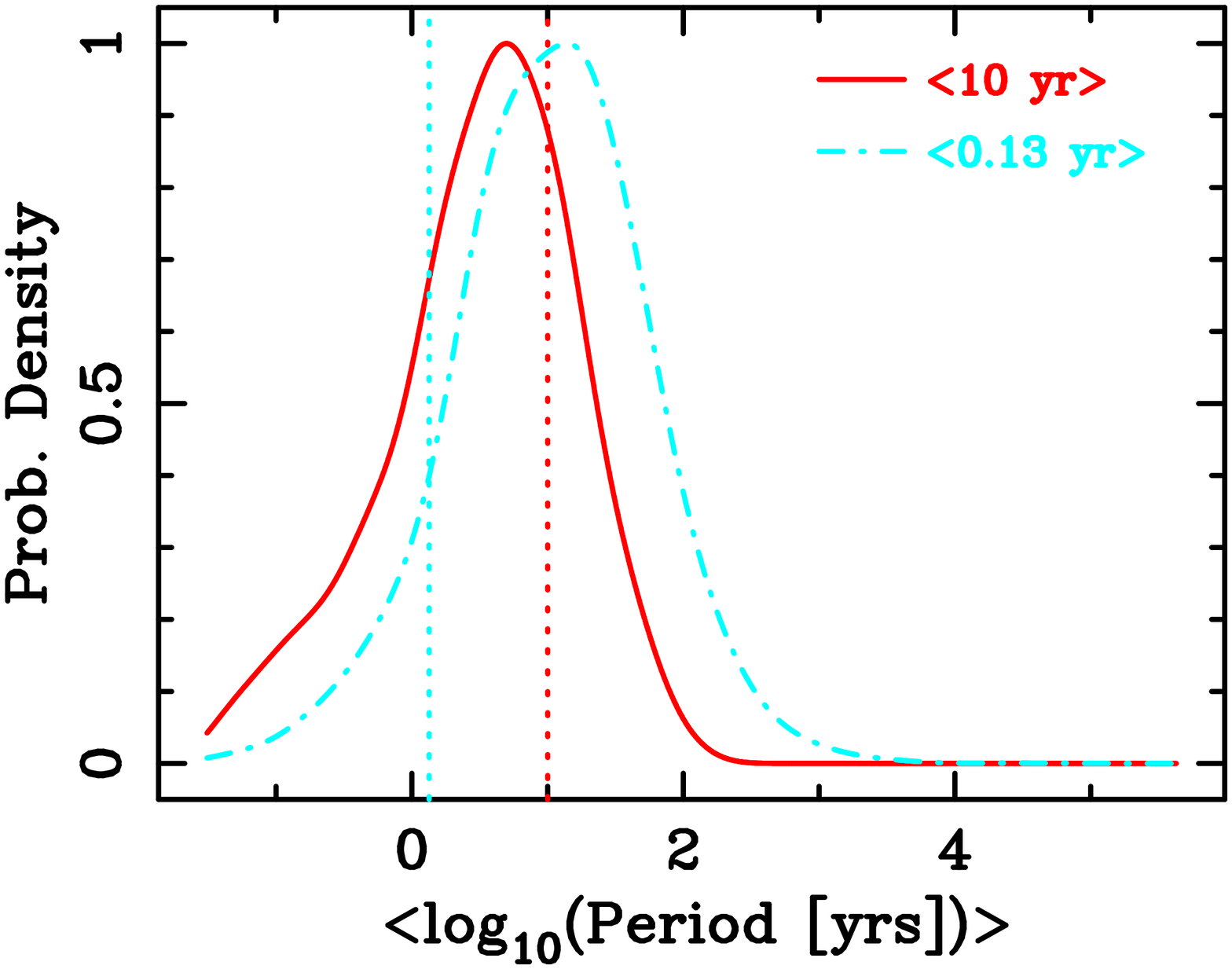}}
		\label{fig:muptest04}
	}
	\subfigure[3.7 km~s$^{-1}$ intrinsic dispersion, 10 year mean period]
	{
		\rotatebox{-90}{\includegraphics[height=0.482\hsize]{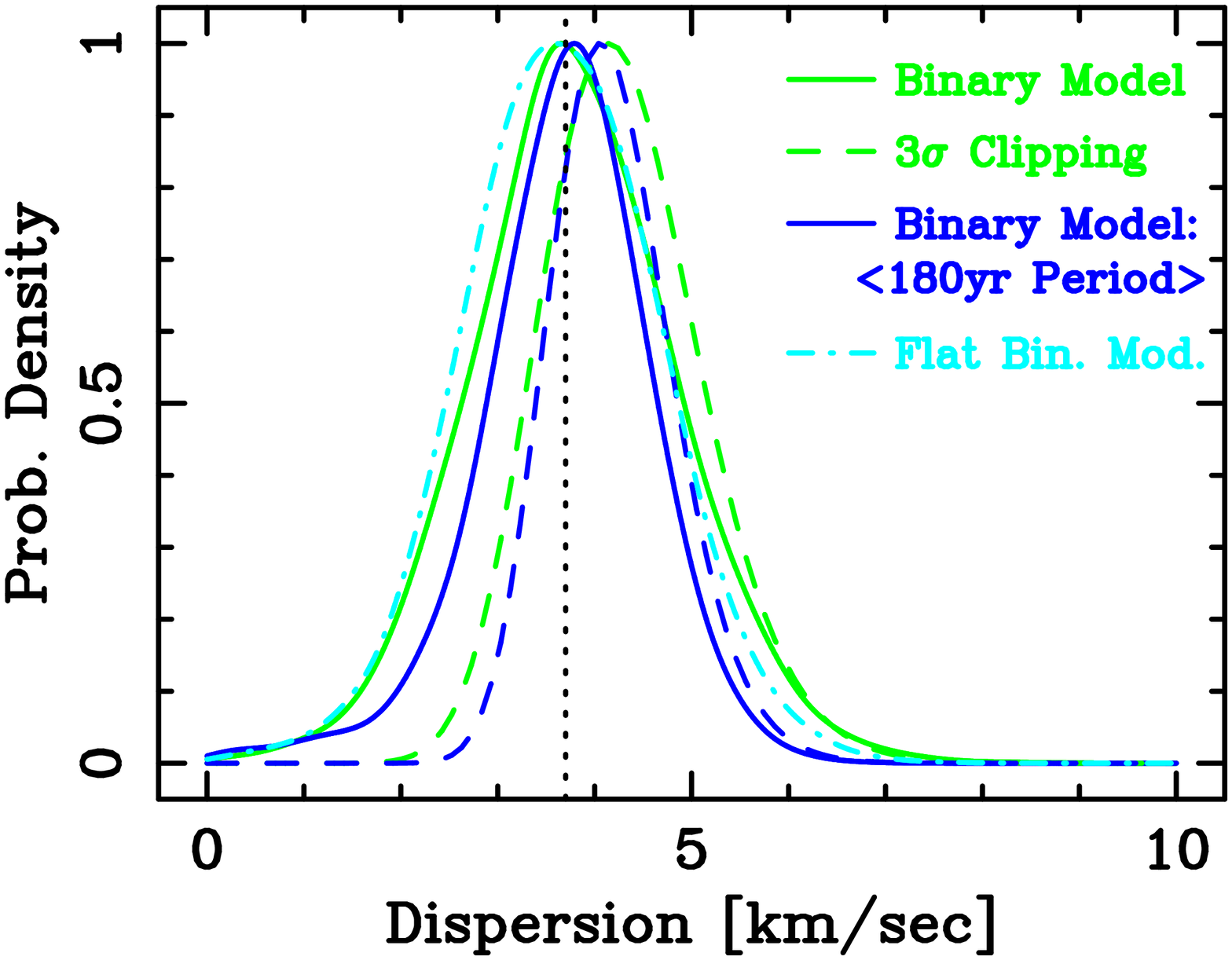}}
		\label{fig:sigtest37}
	}
	\subfigure[3.7 km~s$^{-1}$ intrinsic dispersion, 10 year mean period]
	{
		\rotatebox{-90}{\includegraphics[height=0.482\hsize]{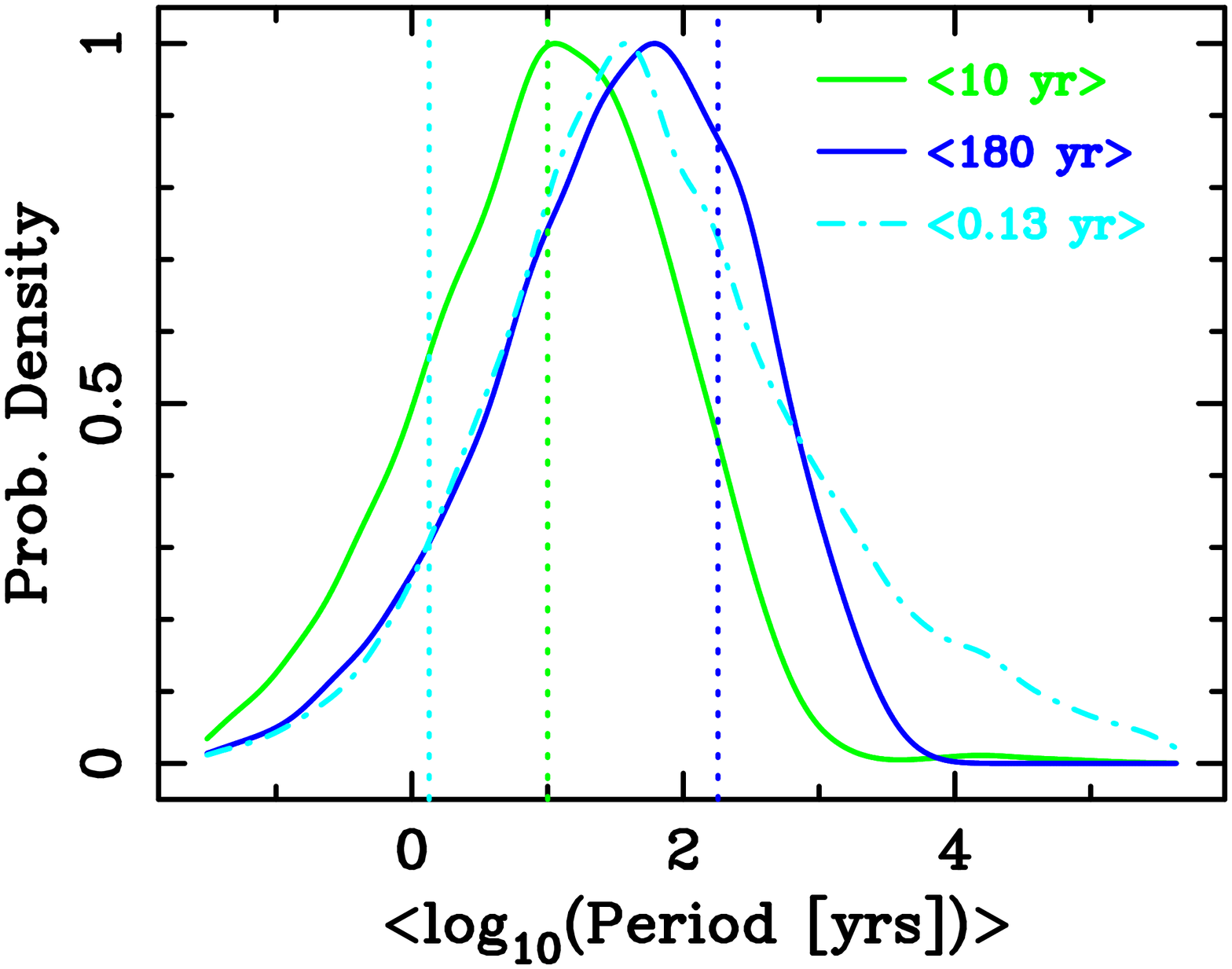}}
		\label{fig:muptest37}
	}
	\caption{\label{fig:simulations}\scriptsize Inferred probability 
distributions of the intrinsic velocity dispersion (panels a and c) and mean binary period 
(panels b and d) for simulated 
Segue 1-like galaxies, using our method of modeling the binary population 
(solid) compared to simply clipping 3$\sigma$ velocity outliers 
(dashed) and then computing the dispersion. Each simulated galaxy
uses the same number of epochs, dates,  velocity errors, and
magnitudes as the actual Segue 1 sample. 
{
This includes foreground stars from the Besancon model for Milky
  Way as outlined in Section~\ref{sec:test}.}   
We 
plot one of the realizations that has a maximum
likelihood velocity dispersion close to 4 km~s$^{-1}$ after  
discarding 3$\sigma$ outliers iteratively.  For the actual intrinsic
dispersion, we choose two cases: 0.4 km~s$^{-1}$  (top panels---a and b) and 3.7 km~s$^{-1}$
(bottom panels---c and d), which is our inferred most probable dispersion of the
actual Segue 1 data set.  The binary population has a mean period of 10
years,  binary fraction of 0.7, and period distribution width
$\sigma_{\log P}=1.5$---consistent with our final results 
for the period distribution and binary fraction of Segue 1 stars.
To infer the binary-corrected dispersion, we marginalize over the systemic 
velocity, binary fraction, mean period, 
foreground parameters ($S$ and $\delta$), and total fraction, whereas for the non-binary 
corrected dispersion we marginalize only over the systemic velocity in addition 
to iteratively discarding 3$\sigma$ velocity outlier stars. It is
clear that our method is able to correctly infer the intrinsic
dispersion and extract the mean period of the binaries (indicated by vertical dotted lines in each panel). 
As a check on the robustness of our methodology, we simulated a
mock data set with orbital periods drawn from a 
distribution flat in logarithm of period (cyan dot-dashed lines).
Employing the same analysis method and assumptions used previously,
the intrinsic dispersion was recovered faithfully for both the 0.4
km~s$^{-1}$ and 3.7 km~s$^{-1}$ cases.
Other realizations show similar behavior, see Section \ref{sec:test} for
more details.}
\end{figure*}

\begin{figure}
\rotatebox{270}{\includegraphics[height=0.964\hsize]{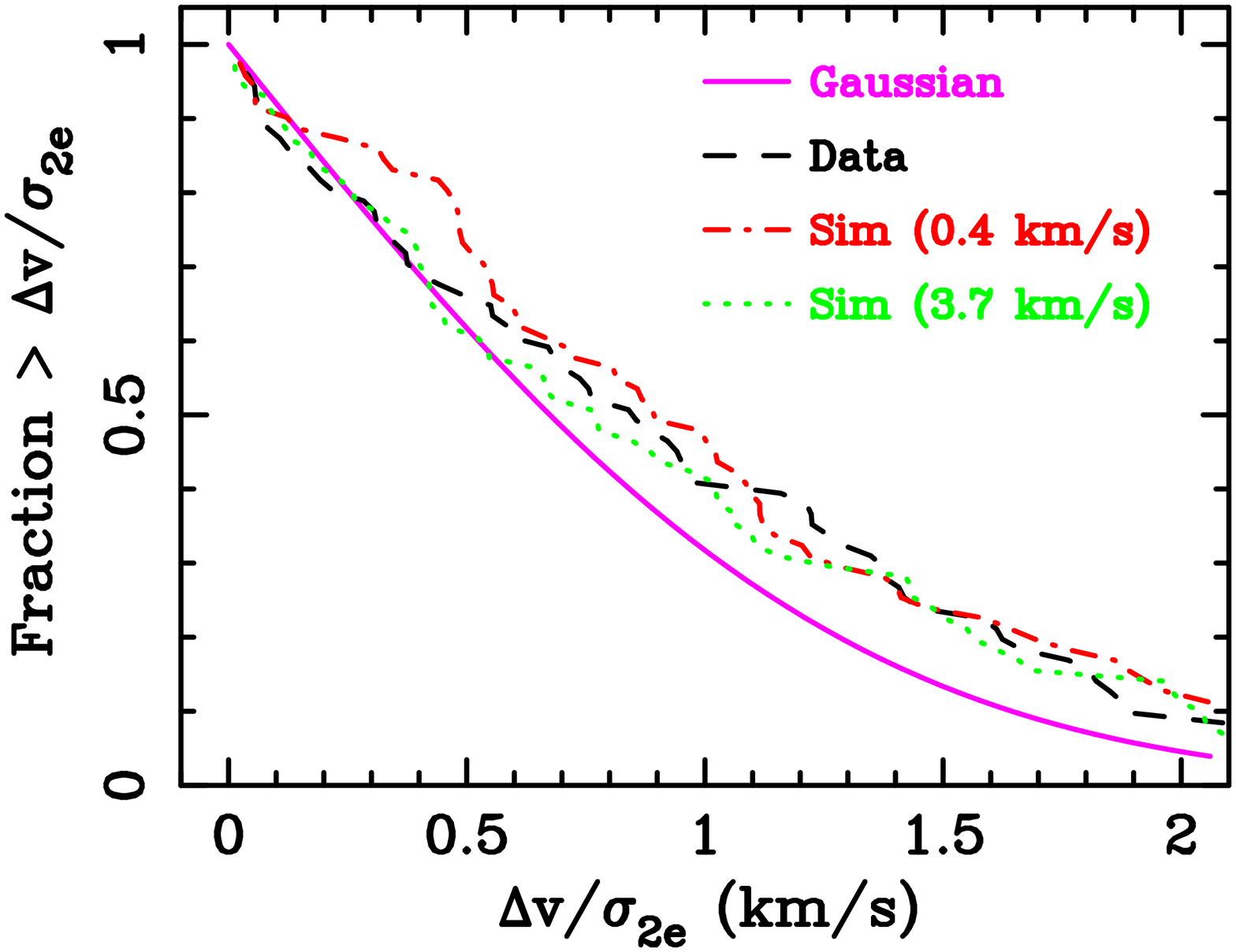}}
\caption{\label{fig:dverr}\footnotesize
Plotted is the fraction of stars with velocity changes (between any
two epochs) greater than a certain threshold defined in units of the
$\sigma^2_{2e}=\sigma_1^2+\sigma_2^2$, where $\sigma_1$ and $\sigma_2$
are the errors of the first and second epoch measurements. The solid
magenta curve is the Gaussian expectation, i.e., no contribution from
binary orbital motion. Simulations with no
binaries (not shown here to retain simplicity of presentation)
straddle the Gaussian expectation and do not show systematic  
positive deviations as large as the data. This is therefore a simple
way to deduce the presence of binaries and it is the reason why we are
able to get a handle on some of the binary properties. We have also
plotted the simulations shown in Figure~\ref{fig:simulations} as the dotted
red and green curves, which show similar behavior as the data and
that is the reason why we are able to constrain their (statistical)
binary properties and recover the intrinsic dispersion.    
}
\end{figure}

\subsection{Test of the multi-epoch binary-correction method\label{sec:test}}

Before writing down our final complete likelihoods for dSph and Milky
Way stars, we provide a {
test of the effectiveness of our binary
  correction method in the presence of foreground stars,
}
which is summarized in
Figure \ref{fig:simulations}. 
We generate a series of mock observations of three Segue 1-like galaxies.
In one set of realizations (illustrated in the upper 
two panels) we assume that the underlying velocity dispersion is
$\sigma = 0.4$ km~s$^{-1}$. Note that because of the extremely low luminosity
of Segue 1 ($L \simeq 400 L_\odot$) even a velocity dispersion as
low as 0.4 km~s$^{-1}$ would imply dark matter, with $M_{1/2}/L_{1/2} \simeq
18$,  using the formula of \citet{wolf2010}. In the other two
realizations (lower panels), we assume that   the intrinsic velocity
dispersion is close to what we infer in the next section  from the
actual dataset, $\sigma = 3.7$ km~s$^{-1}$.  
We assume that the galaxies have binary star populations of the type
that could conceivably hinder our ability to infer an 
intrinsic velocity dispersion, with $B=0.7$, a mean  period of $P=10$
yr, a period spread of $\sigma_{\log P}=1.5$, 
ellipticity and mass
fraction distributions as described in the earlier section. 
{
We also include a  distribution of foreground stars drawn from a
  Besancon model displaced 209 km~s$^{-1}$ in mean velocity (see
  Figure~\ref{fig:binneddata}) from the mock Segue 1 galaxy.}    
In the  simulations with $\sigma = 3.7$ km~s$^{-1}$, we also consider a case
with a 180 year mean period, which is consistent with the solar
neighborhood value.   
In addition, we simulated a set of mock
  observations with periods drawn from a distribution flat in
  $\log P$ for both the $0.4$ km~s$^{-1}$ and $3.7$ km~s$^{-1}$
  cases. We analyze this set in the same way as others by
  fitting it with a log-normal distribution in period. 
Each mock galaxy consists of 69 member stars and they are
``observed'' once or  multiple times in  exact correspondence with our
real Segue 1 member sample  (Paper I, Section 3.1).  
Each mock
  data set also contains 
  109 nonmember foreground stars  randomly selected from the
  Besancon model. 
For each star, velocities are generated using the measurement errors, 
dates, and magnitudes from the measured stars in the Segue 1
sample. In each panel,  the dashed curve shows the result of inferring
the intrinsic dispersion based on the common procedure of  
discarding 3$\sigma$ outliers iteratively {
from the member star sample
}  
(labeled as 3$\sigma$ 
clipping).  The solid lines in each panel show the  probability
distributions of intrinsic dispersions using our Bayesian analysis of
multi-epoch data described in the previous sections. 
The dot-dashed curves show the probability distributions resulting
from our full Bayesian analysis on a mock data set simulated with a
flat $\log P$ distribution. 

The top panel of Figure~\ref{fig:simulations} illustrates that even
when binary orbital motion accounts for most of the {\em observed}
dispersion {
in the presence of significant foreground contamination}, 
our method is able to  extract the intrinsic dispersion
faithfully.  Moreover, we are able to  recover the mean binary period,
including cases where the  intrinsic dispersion is fairly high (lower
panel) and even when $P=180$ yr.  Although periods longer than a few
years are not directly observable in the time-frame of our mock
observations (1--2 years), our method extrapolates from the period
distribution at shorter periods under the assumption of a log-normal
period distribution. Therefore, the mean period can still be inferred
in this case, inasmuch as the assumption of a log-normal period
distribution holds. 
 More impressively, our method is able to 
  recover the correct intrinsic dispersions even when the underlying
  period distribution is different from log-normal.

Though only typical simulation results are shown in
Figure~\ref{fig:simulations}, we have applied  our method to several
mock galaxies within each category described above.  In the case of a
0.4 km~s$^{-1}$ intrinsic dispersion, our inferred dispersions are
consistently more accurate than the $3\sigma$ clipping result, and
each distribution exhibits only a  small probability of having a dispersion
greater than $\sim$3 km~s$^{-1}$. Furthermore, the most probable inferred
dispersion corresponding to the peaks  in Figure~\ref{fig:sigtest04} is
smaller than 2 km~s$^{-1}$ for every simulation we ran. These  results show
that binaries account for most of the {\em observed} dispersion in
this set of simulations, and that our method is able  to extract the
intrinsic dispersion faithfully.  In  the case of 3.7 km~s$^{-1}$ intrinsic
dispersion (Figure~\ref{fig:sigtest37}), while  some of the posteriors
do have tails going to zero dispersion, the probability  in the region
between 0--1 km~s$^{-1}$ dispersions is quite small, confirming  that
binaries are unlikely to account for most of the 4 km~s$^{-1}$ observed
dispersion.
 In all cases where the periods in the mock data were drawn from
  a log-normal period distribution, we are also able to recover the mean binary
period to within about 1$\sigma$  (as the example in
figs.~\ref{fig:muptest04} and \ref{fig:muptest37} illustrates).  

It is worth noting that there is a small possibility of failure in
extracting the correct intrinsic dispersion and mean period. This
failure can arise in the following way. If the simulated data set has
a few outliers in velocity changes but the data are otherwise roughly 
consistent with the measurement errors, then the method will try to fit
the outliers with a mean period smaller than the correct one. However,
the small number of significant velocity changes will force the binary
fraction to be small. The shorter mean period will force the intrinsic
dispersion to be smaller (than the true dispersion), while the
smaller binary fraction will reduce the contribution of binary orbital
motion and hence increase the inferred intrinsic dispersion. These
effects go in opposite directions, but the net effect could be to
underestimate the  binary correction. To test for this
possibility, we plot in Figure~\ref{fig:dverr} the fraction of velocity
changes greater than a threshold and compare to the expectation from
just measurement errors. The data (shown in dashed black) 
exceed the Gaussian (measurement error) expectation and hence
provide visceral proof of the presence of binary stars. There is a 
systematic positive deviation that is not consistent with changes 
introduced by purely measurement errors. These
deviations allow us to deduce the effect of the binary orbital motion
on the measured dispersion. The fact that our data set does not show 
signatures of just few outliers (i.e., it shows a systematic positive 
deviation from the Gaussian expectation) also assures us that it is not
prone to the failure mode described above.  

These simulation results provide ample reasons for confidence in our 
technique. We will proceed to apply this technique to the real Segue 1
data set in Section  \ref{sec:dispersion_results}.  Before doing so,
we add the final piece to our  likelihood, which will allow for a
self-consistent Bayesian treatment of both  binary orbital motion and
membership probabilities. 

\subsection{Likelihood for dwarf spheroidal and Milky Way stars}

Suppose that over a certain region of the sky containing the dwarf galaxy 
sample, a fraction $F$ of the stars are members. 
 We can  then write the likelihood in terms of $F$ as
\begin{eqnarray}
&& \mathcal{L}(v_i|e_i,t_i,M;F,B,\sigma,\mu) ~ = ~ (1-F)\mathcal{L}_{\mathrm{MW}}(v_i|e_i) \nonumber \\
& & ~ ~ + ~ F \mathcal{L}_G(v_i|e_i,t_i,M;B,\sigma,\mu) \, ,
\label{eq:like}
\end{eqnarray}
where $\mathcal{L}_G$ is the galaxy likelihood given by Equation~(\ref{eq:full_binary_likelihood}) 
and $\mathcal{L}_{\mathrm{MW}}$ is the Milky Way likelihood.
Since the majority of nonmember stars were not singled out for repeat 
measurements, we do not directly model the binary population of the Milky Way; 
instead, the effect of binaries and other uncertainties in the Milky Way 
velocity distribution are accounted for by the translation and stretch 
parameters $\delta$ and $S$ (as discussed above in Equation~(7)).  For compactness, we 
have suppressed the $\delta$ and $S$ dependence in Equation~(\ref{eq:like}) and have 
also suppressed the metallicity and position components of the likelihood, 
which are, however, still included in our marginalization.

The Milky Way likelihood $\mathcal{L}_{\mathrm{MW}}$ can be written as
\begin{equation} \mathcal{L}_{\mathrm{MW}}(v_i|e_i) = \int 
P_{\mathrm{MW}}(v_i|v_{\mathrm{cm}},e_i)P_{\mathrm{MW}}(v_{\mathrm{cm}})dv_{\mathrm{cm}} \, , \end{equation}
where $P_{\mathrm{MW}}(v_{\mathrm{cm}})$ is obtained from the Besancon model, and the 
distribution $P_{\mathrm{MW}}(v_i|v_{\mathrm{cm}},e_i)$ is equivalent to that of 
Equation~(\ref{eq:v_given_vcm_like}) with $B=0$. Plugging in 
Equations~(\ref{eq:v_given_vcm_like}) and (\ref{eq:full_binary_likelihood}), we arrive 
at
\begin{eqnarray}
&& \mathcal{L}(v_i|e_i,t_i,M;F,B,\sigma,\mu) ~ \propto ~ (1-F)\mathcal{L}_{\mathrm{MW}}(\langle v \rangle|e_m) \nonumber \\
& & ~ ~ + ~ F\left\{(1-B)\frac{e^{-\frac{(\langle v\rangle - \mu)^2}{2(\sigma^2 + e_m^2)}}}{\sqrt{2\pi(\sigma^2 + e_m^2)}} + BJ(\sigma,\mu)\right\} \, .
\label{eq:full_monty_likelihood}
\end{eqnarray}
Again, we have absorbed the normalizing factor $\mathcal{N}$ into the definition of 
$J(\sigma,\mu)$ (given in Equation~(\ref{eq:i_integral}))  and
\begin{equation}
\mathcal{L}_{\mathrm{MW}}(\langle v \rangle|e_m) = \int \frac{e^{-\frac{(v_{\mathrm{cm}} - \langle v\rangle)^2}{2 e_m^2}}}{\sqrt{2\pi e_m^2}} P_{\mathrm{MW}}(v_{\mathrm{cm}}) dv_{\mathrm{cm}}.
\end{equation}

Each term in Equation~\ref{eq:full_monty_likelihood} gives the relative 
likelihood of being a Milky Way, Segue 1 single, and Segue 1 binary star respectively. Again, 
we emphasize that our full likelihood also includes metallicity and position 
information to help determine membership of each star. To accomplish this, we 
multiply each term in Equation~(\ref{eq:full_monty_likelihood}) by the corresponding 
likelihoods in metallicity and position (Equation~(\ref{eq:likelihood_vwr})), the form 
of which has already been discussed in Section \ref{sec:membership_method}. In 
addition to the dispersion and systemic velocity of the dwarf galaxy, there are 
9 model parameters to help determine membership 
(Equation~(\ref{eq:lots_of_parameters})) and 3 binary parameters ($B$, $\mu_{\log P}$, 
$\sigma_{\log P}$), for a total of 14 model parameters in the likelihood.

\section{Velocity dispersion of Segue 1}\label{sec:dispersion_results}

\begin{figure*}
\rotatebox{270}{\includegraphics[height=0.482\hsize]{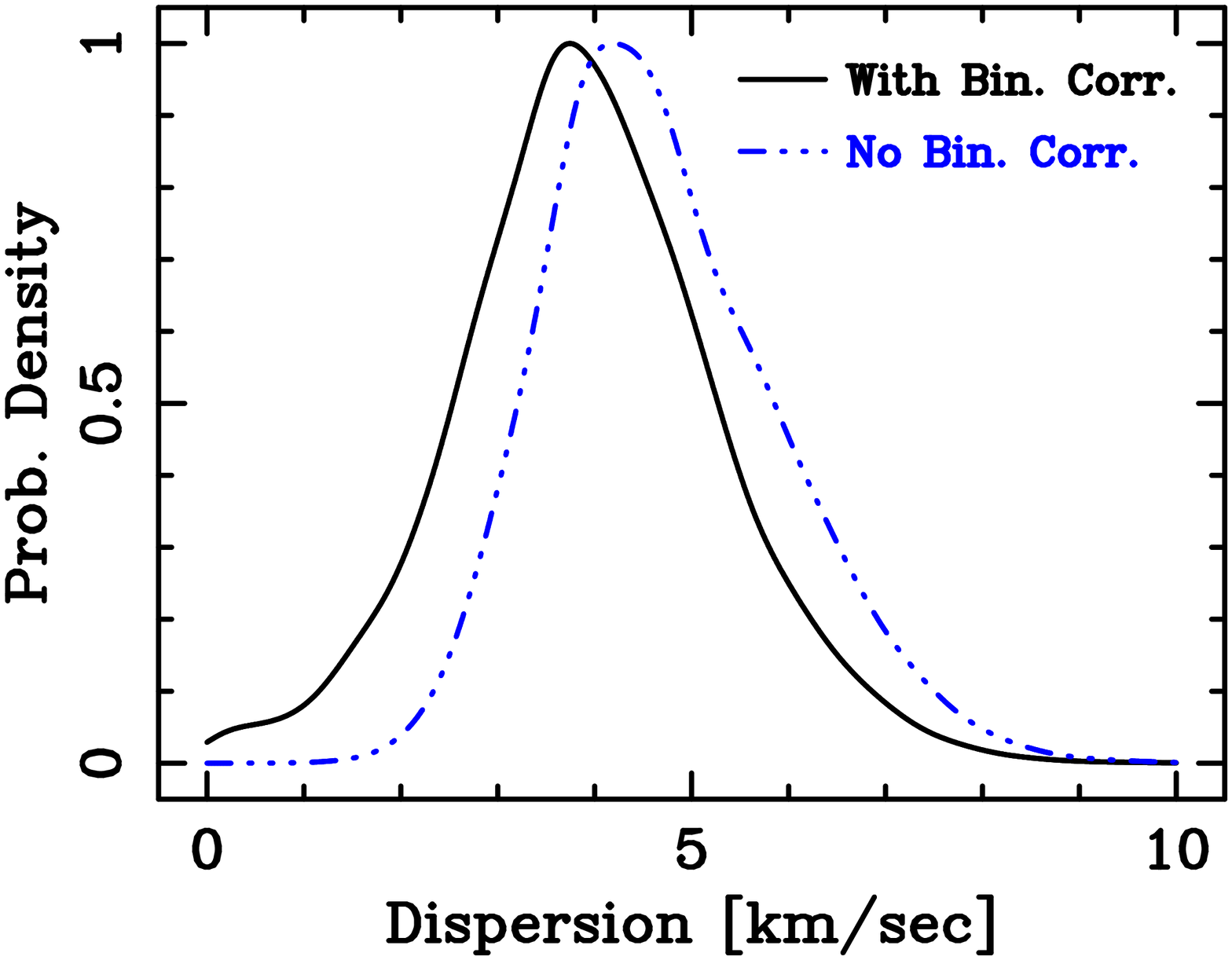}}
\rotatebox{270}{\includegraphics[height=0.482\hsize]{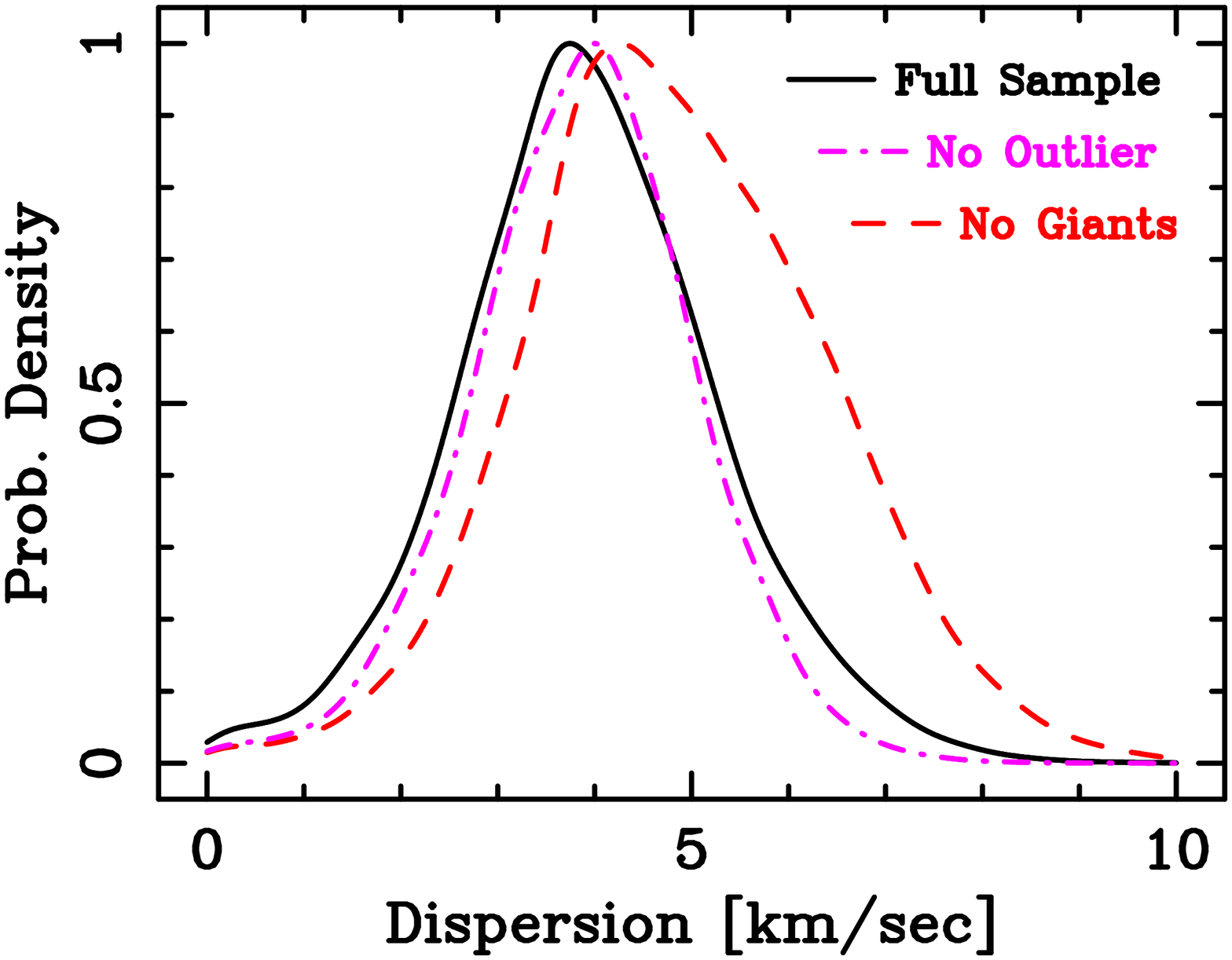}}
\caption{\label{fig:dispersion}\footnotesize Inferred probability 
density of the velocity dispersion of Segue 1. 
Left:  comparing the probability density with (solid black line) and without (triple-dot-dashed 
blue line) the correction due to binary motion, we see that correcting for 
binaries results in a lower inferred dispersion and gives rise to a tail at low 
velocities, due mainly to short-period binaries (Section 
\ref{sec:mean_period_results}). 
Right: note that excluding the star 
SDSSJ100704.35+160459.4, which is a 6$\sigma$ velocity outlier with a 
substantial membership probability, does not have a significant impact on the 
inferred dispersion (dash-dotted cyan line) since its possible membership and 
binarity are treated statistically (see paragraph 2 of Section \ref{sec:dispersion_results}).  
Exclusion of the red giants biases the 
probability distribution (dashed red line) to higher dispersion values; this is 
primarily due to their smaller measurement errors which give them a large 
relative weight in determining the velocity dispersion despite the small number 
of probable members (six RGB stars in total).
}
\end{figure*}

\begin{figure*}[h!]
	\subfigure[intrinsic dispersion vs. mean log of period]
	{
		\rotatebox{0}{\includegraphics[height=0.482\hsize]{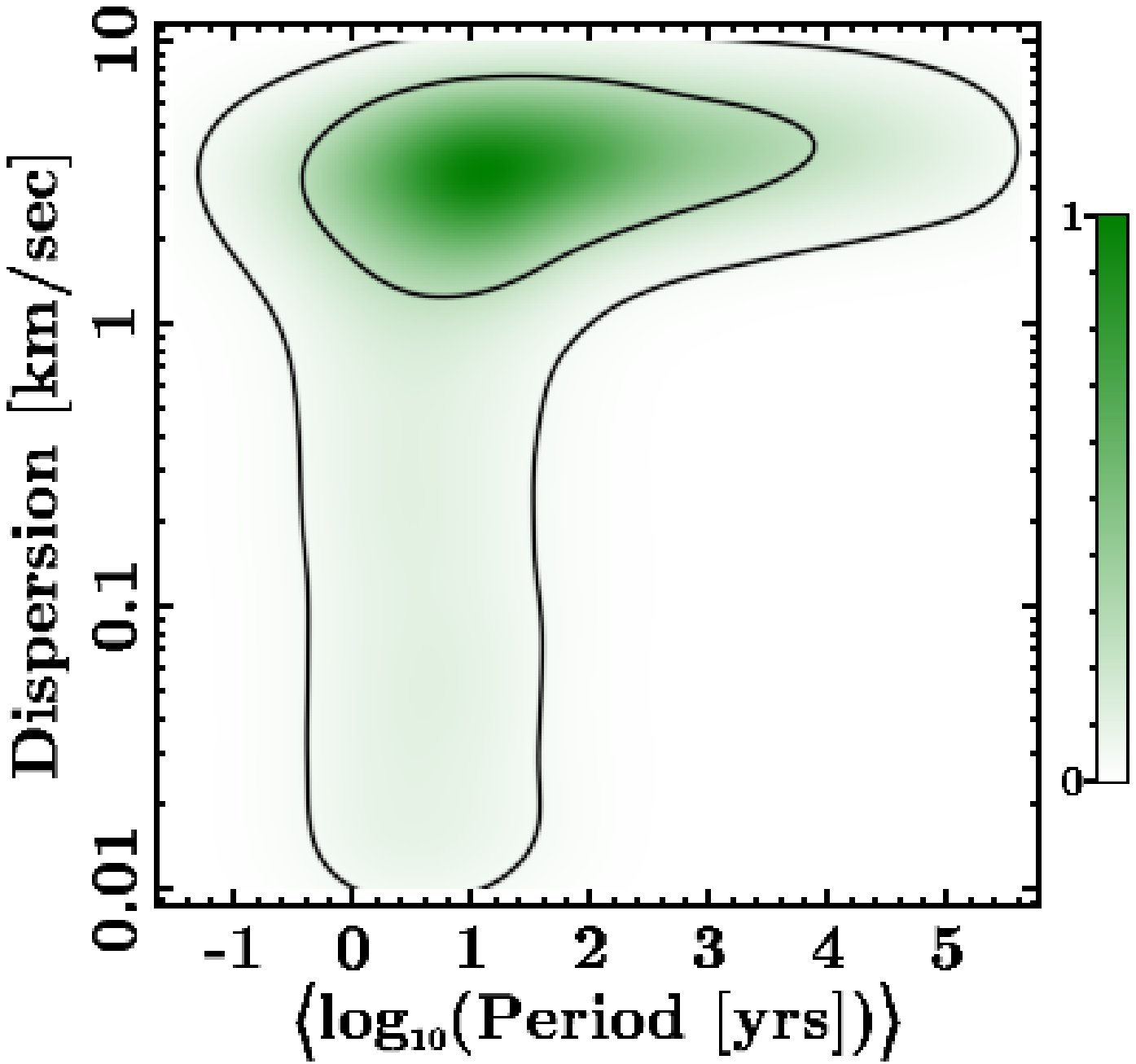}}
 		\label{fig:sigvsmup}
	}
	\subfigure[width of period distribution vs. mean log of period]
	{
		\rotatebox{0}{\includegraphics[height=0.482\hsize]{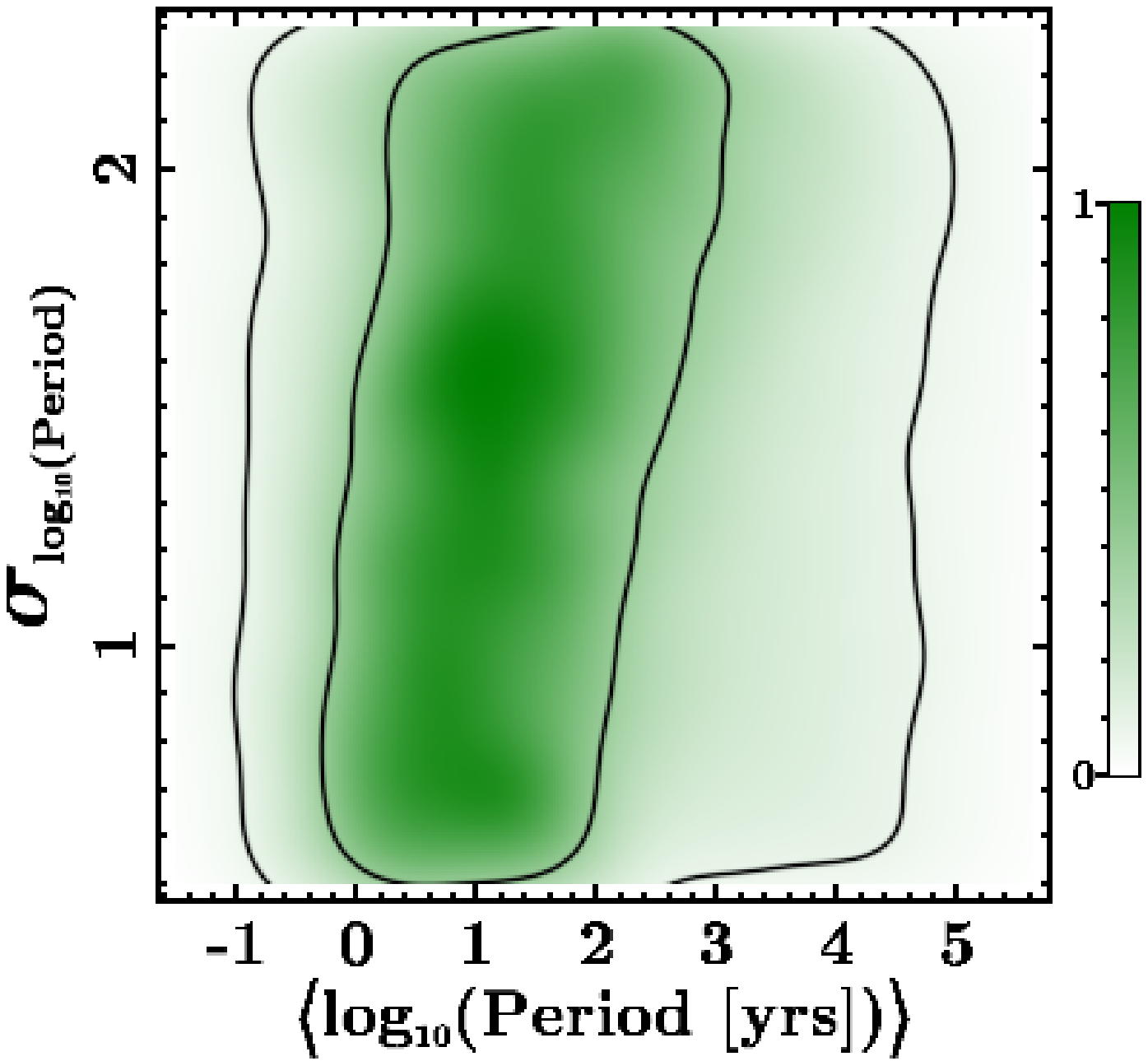}}
		\label{fig:sigpvsmup}
	}
	\subfigure[Mean Ca width vs. mean log of period]
	{
		\rotatebox{0}{\includegraphics[height=0.482\hsize]{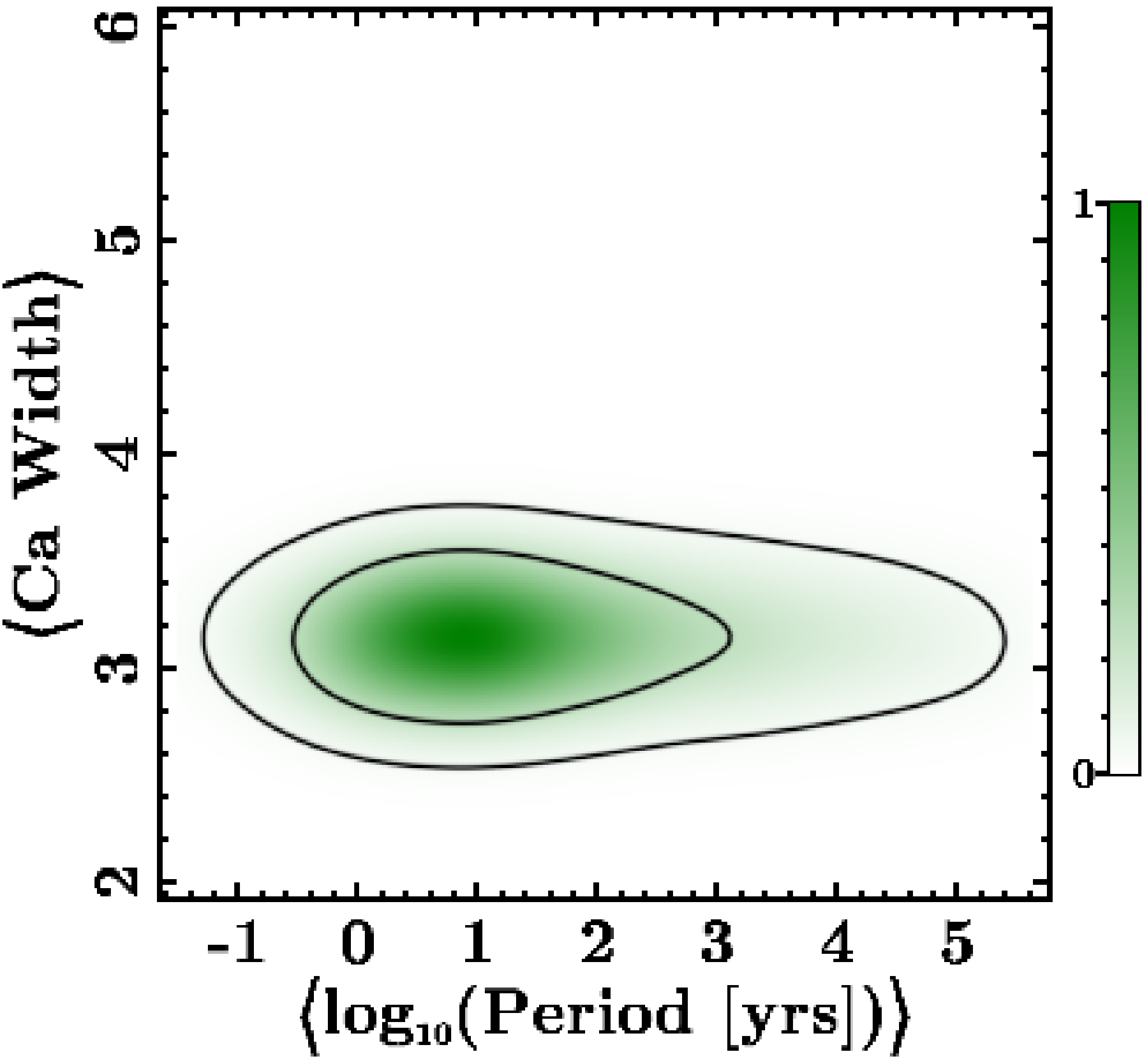}}
		\label{fig:metvsmup}
	}
	\subfigure[binary fraction vs. mean log of period]
	{
		\rotatebox{0}{\includegraphics[height=0.482\hsize]{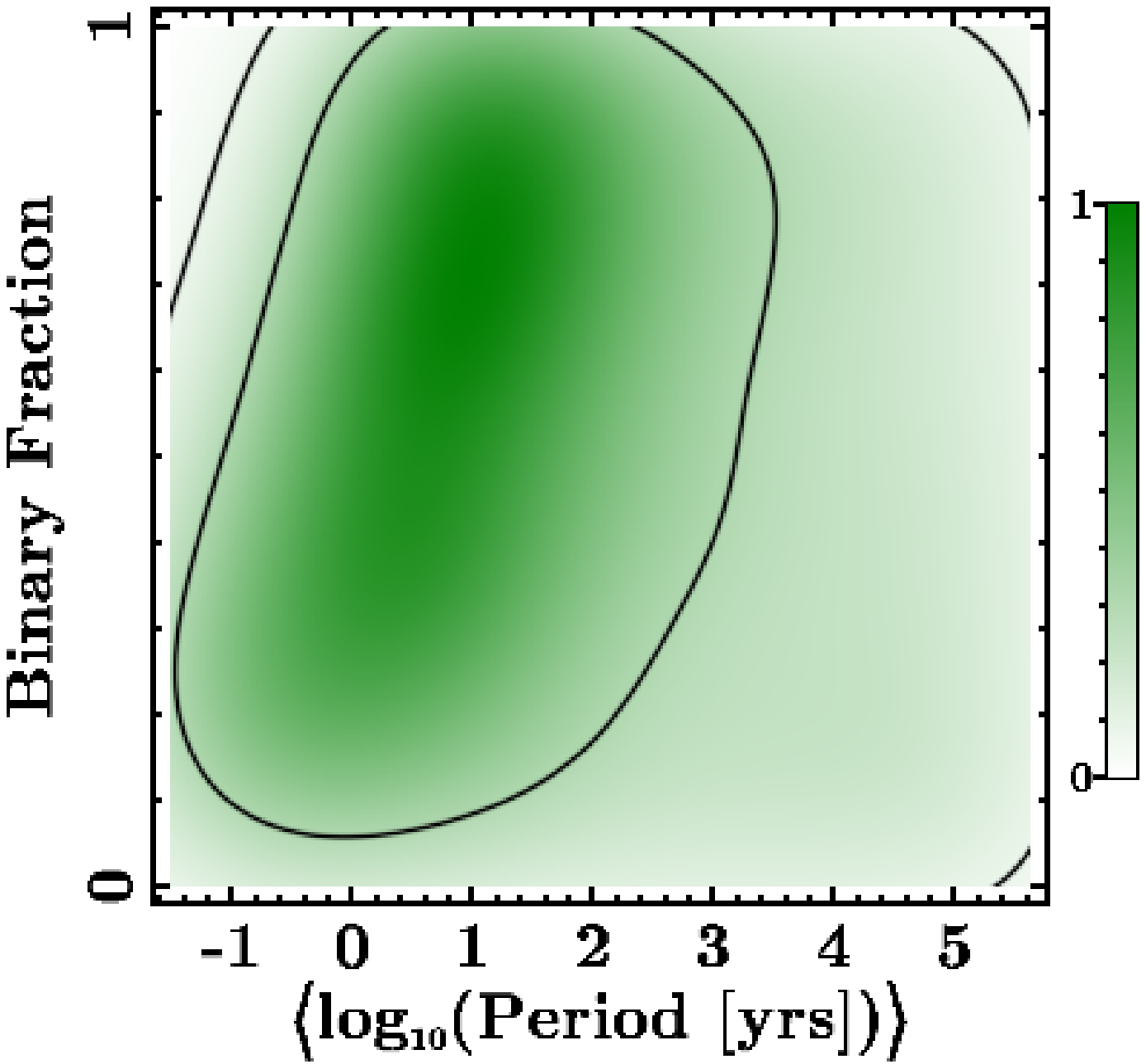}}
		\label{fig:bvsmup}
	}
	\caption{\label{fig:mupvs}\footnotesize Joint posterior probability 
distributions of: 
(a) intrinsic dispersion vs. mean
log-period of Segue 1 binary population, 
(b) width of period distribution vs. mean log-period, 
(c) mean Ca width of the member stars vs. mean log-period, and
(d) binary fraction vs. mean log-period. Inner and outer contours 
surround the region containing 68\% and 95\% of the total probability, 
respectively.}
\end{figure*}

Using the method outlined in the previous sections, we infer a posterior 
probability distribution for the intrinsic velocity dispersion of Segue 1 by 
marginalizing over all the other parameters via a nested sampling routine \citep{skilling04, feroz2009}.  In 
the left panel of Figure~\ref{fig:dispersion}, the inferred probability distribution of the 
galaxy's velocity dispersion is plotted with (solid) and without (triple-dot-dashed) 
correcting for binaries.  We see that correcting for binaries lowers the 
inferred dispersion slightly and gives rise to a small but non-zero 
probability of an intrinsic dispersion smaller than 1.5 km~s$^{-1}$.  Using our full 
sample, the binary-corrected velocity dispersion is $3.7^{+1.4}_{-1.1}$ at
1$\sigma$.  We find a $\sim 3.5\%$ probability of a dispersion smaller than 1.5 
km~s$^{-1}$, and $\sim 1.7\%$ for dispersions $<$ 1 km~s$^{-1}$.  Although the low-dispersion 
tail in the probability distribution is small, it does extend all the way to 
zero velocity dispersion. As we will show in Section 
\ref{sec:mean_period_results}, this is due mainly to the possibility of 
binary populations with a  high binary fraction and a mean period shorter 
than 10 years (see Figure~\ref{fig:sigvsmup}).

To test how sensitive our results are to individual velocity outlier stars, 
the right panel of Figure~\ref{fig:dispersion} plots the inferred dispersion if the star 
SDSSJ100704.35+160459.4 is excluded from the sample.  This star is a 6$\sigma$ 
velocity outlier that nevertheless has a substantial membership probability 
($\left<p\right>=0.49$) due to its proximity to the projected center of Segue 1.  
The inferred maximum likelihood dispersion using 
the expectation maximization algorithm of \citet[which is not corrected for
binaries]{walker11-09} decreases from 5.5 km~s$^{-1}$ to 3.9 km~s$^{-1}$ when
SDSSJ100704.35+160459.4 is removed from the sample (Paper I).
We find that excluding SDSSJ100704.35+160459.4 does not have a significant 
effect on the general properties of the dispersion probability 
distribution---the spread, $\approx$ 4 km~s$^{-1}$ peak, and low-velocity tail 
features are largely unaffected.  This is partly because its membership is 
treated in a statistical sense, and also because if the star is a member of 
Segue 1, the implied probability of being a binary is quite high 
($\left<p_b\right>=0.89$).  
On the 
other hand, excluding the giants from the sample does bias the result to higher 
dispersion values.  This is primarily due to the smaller measurement errors in 
the red giant population which give them a high relative weight in determining 
the velocity dispersion despite their small numbers (six giant branch stars in total). 

We have also investigated the intrinsic dispersion of these giant branch stars. 
Using the same method outlined above but removing the main-sequence stars
that are identified as Segue 1 members in Paper I, we find that the
intrinsic dispersion is $2.0^{+3.1}_{-1.7}$ km~s$^{-1}$.  This is consistent 
with the dispersion obtained from the full sample.  The large error bars are
due to small number of Segue 1 members as compared to the Milky Way stars in the sample.
Using the less statistically rigorous method of assuming these six giant branch
stars are definite members and not including any other stars 
(or a second Milky Way distribution in the likelihood), 
we obtain a dispersion of $1.7^{+1.2}_{-1.3}$ km~s$^{-1}$.
While it is in principle possible that the giants and the main
sequence stars could trace two kinematically distinct populations,
this scenario is physically very unlikely because the ages and
masses of the giants should be effectively identical to those of their
less-evolved counterparts.  However, the small number of stars on the
RGB precludes us from conclusively ruling out such an
occurrence.

\begin{figure}[t!]
\rotatebox{270}{\includegraphics[height=0.964\hsize]{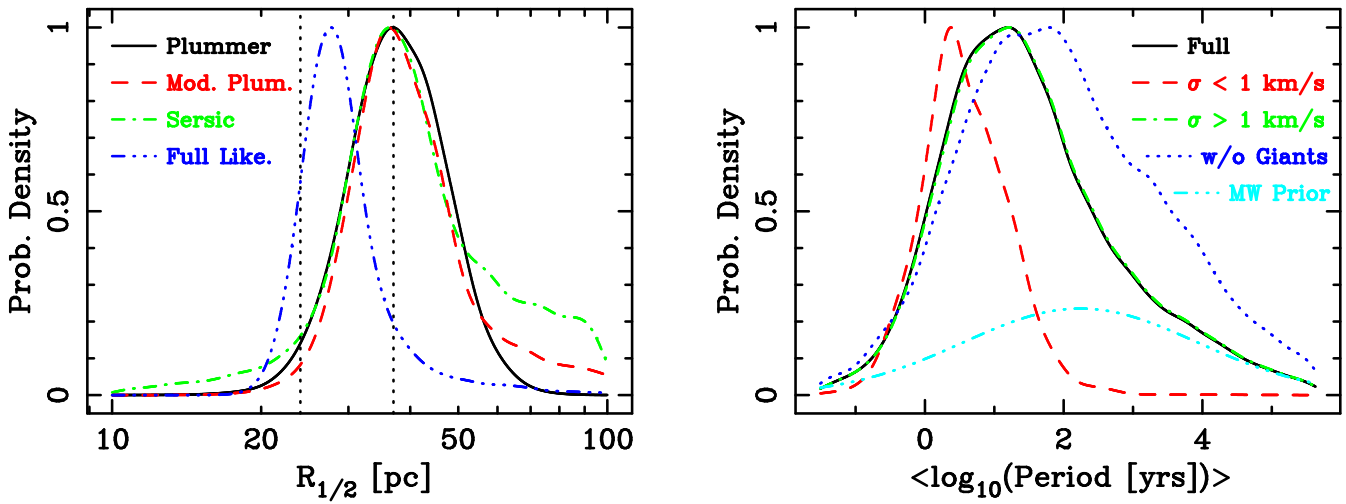}}
\caption{\label{fig:r12}\footnotesize 
Probability density of the projected (two-dimensional) radius containing half the member stars of Segue 1.
 The wider distributions 
were computed using our conditional likelihood (Equation~(\ref{eq:condlike})) assuming  a Plummer model (solid black), a
modified Plummer model (dash-dotted green), and a Sersic model (dashed red) for the stellar density
profile.   The triple-dot-dashed blue line shows that the probability density
is further constrained when we include the full position information in the likelihood ($L(v,w,r)$).  The
vertical lines bracket the 68\% confidence region of the best photometric determination 
of $R_{1/2}$ \citep{martin2008}.  
}
\end{figure}

One particularly robust constraint from our analysis
concerns the stellar number density profile. Figure \ref{fig:r12}
shows the inferred probability density of the projected
(two-dimensional) radius that contains half of the member stars of
Segue 1 ($R_{1/2}$).    Three of the probability densities plotted
use our standard conditional likelihood ($L(v,w \vert R)$, Equation~(\ref{eq:condlike})) and
explore how our derived $R_{1/2}$ constraints depend on the
assumed stellar density profile shape.  The solid, dash-dotted, and dotted
lines, respectively, are derived using a
Plummer model, a Sersic model, and 
a modified Plummer model (see Equation~(\ref{eq:modplum}))
where the outer slope ($\alpha$) is marginalized from 3 to 10. Regardless of the 
assumed stellar density profile, $R_{1/2}$ is typically constrained
to be $30$--$50$ pc.  The triple-dot-dashed line shows the probability density
that results when we include the position
information directly in the likelihood using Eq.~\ref{eq:lgalmw}.  In
this case, the constraints 
on $R_{1/2}$ get even tighter, with $R_{1/2} = 28^{+5}_{-4}$ pc, which is in 
good agreement with the best photometric determination of $R_{1/2} = 
29^{+8}_{-5}$ \citep[][1$\sigma$ range 
indicated with vertical dotted lines]{martin2008}. We emphasize
 that we have not used any prior on the light distribution of Segue 1
 from photometry in our analysis. The $R_{1/2}$ determinations shown
 in Figure \ref{fig:r12} are derived entirely from our complete
 kinematic sample. 

\begin{figure}
\rotatebox{270}{\includegraphics[height=0.964\hsize]{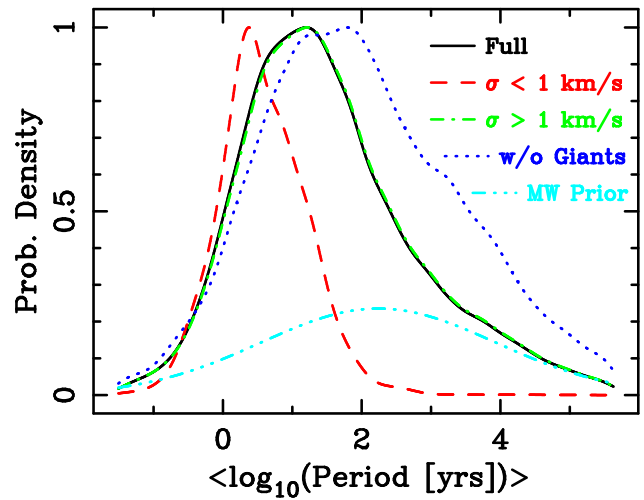}}
\caption{\label{fig:mup}\footnotesize Probability density of the mean 
log-period of Segue 1's binary population (solid curve). For comparison we plot 
our fiducial prior on the mean period (triple-dot dashed curve), which is determined by 
the requirement that a large number of binary populations drawn from this prior 
superimpose to form the log-normal period distribution of field binaries 
observed in the solar neighborhood, which have a mean period of 180 years. We 
see that the data suggest that the mean period of Segue 1 may be significantly 
shorter than that of field binaries, with a most probable inferred mean period 
of $\approx$ 10 years. The dashed (dot-dashed) curve is the distribution in the 
parameter space where the inferred dispersion is smaller (larger) than 1 km~s$^{-1}$.}
\end{figure}
\begin{figure*}
\rotatebox{270}{\includegraphics[height=0.482\hsize]{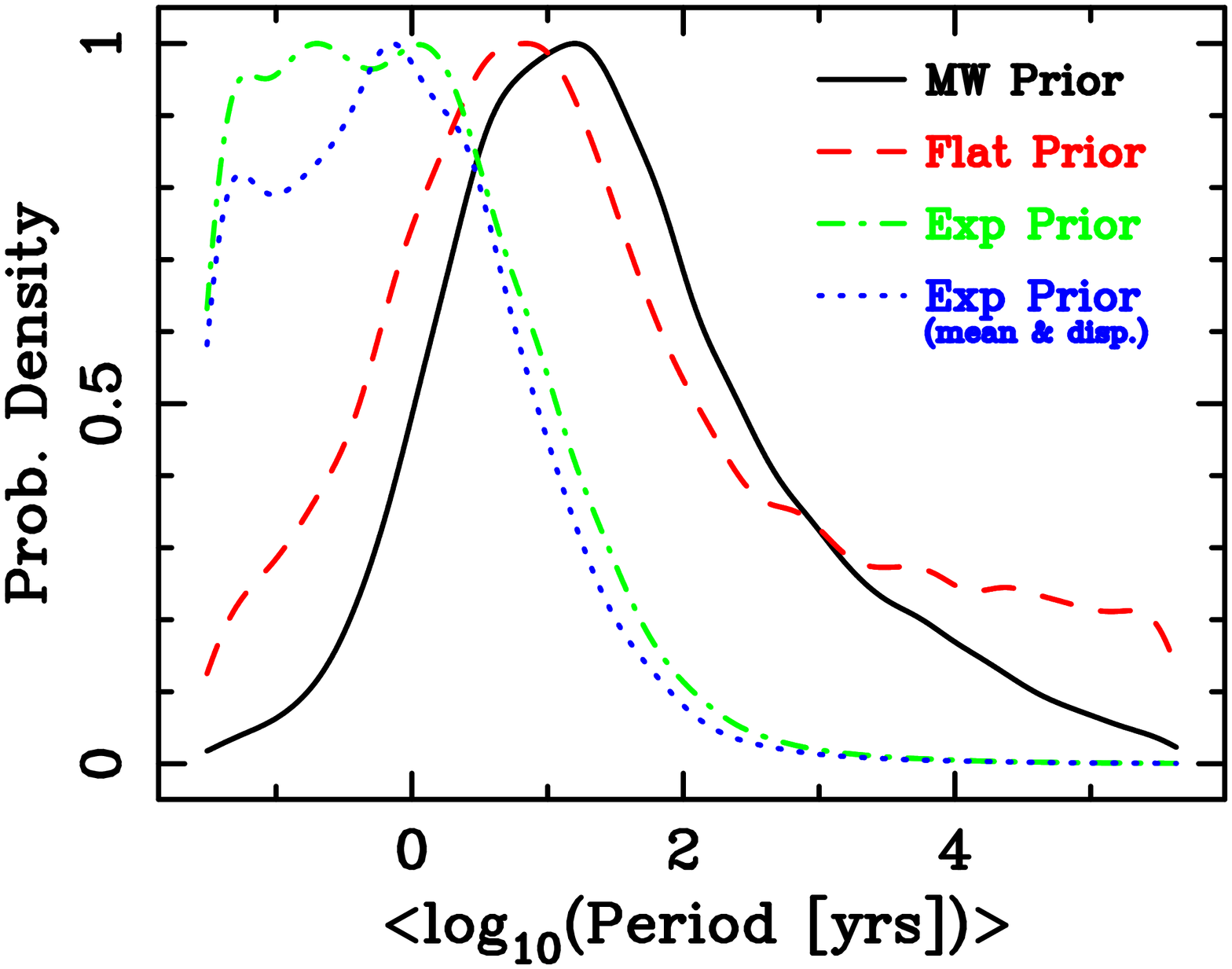}}
\rotatebox{270}{\includegraphics[height=0.482\hsize]{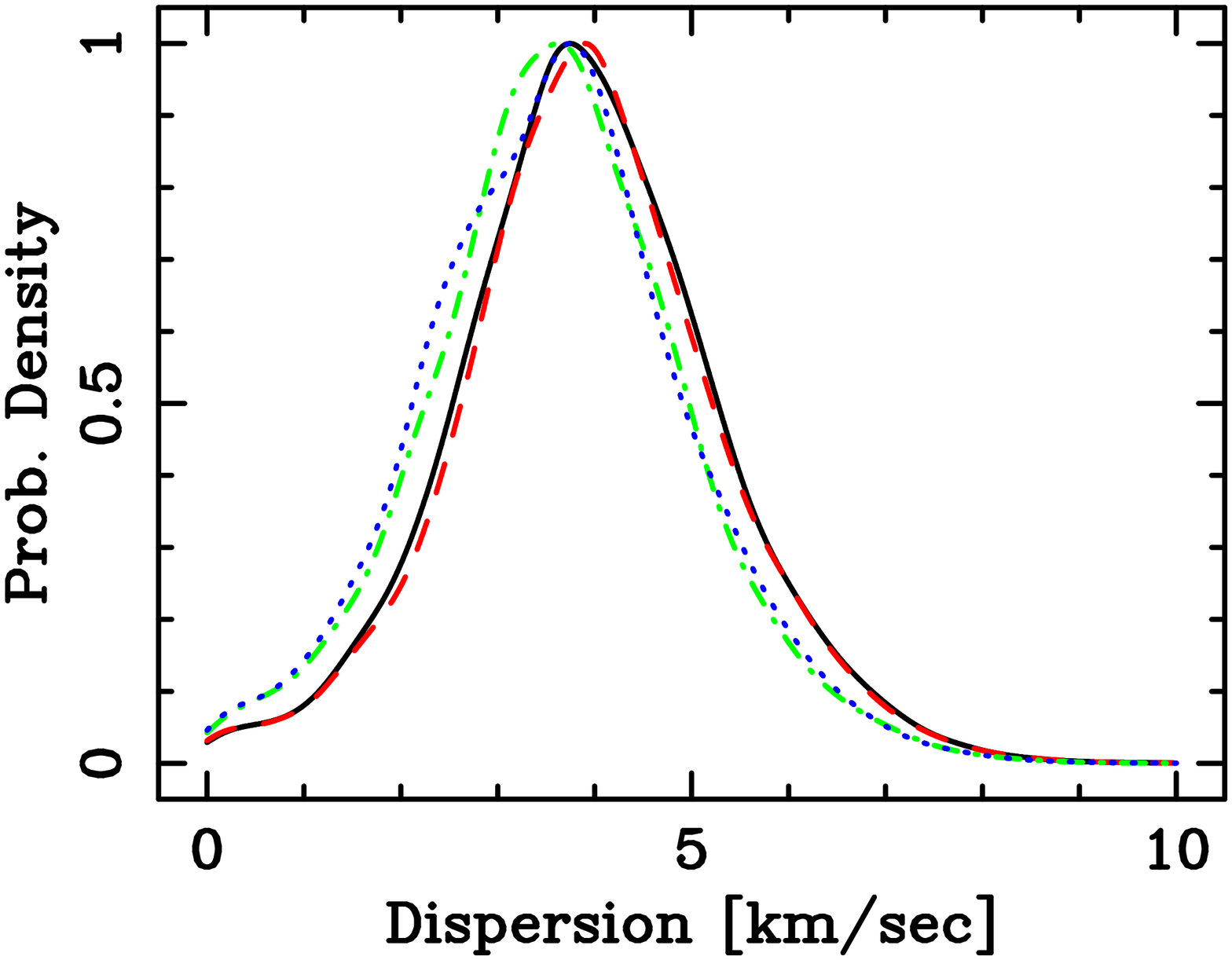}}
\caption{\label{fig:binprior}\footnotesize Plotted are the probability 
densities of the mean orbital period and the corresponding dispersion assuming 
various priors on the binary distribution parameters.  Although the period 
distribution is not well constrained by data, we find that the dispersion 
probability density is surprisingly robust to the shape of the period 
distribution.  Here, we compare our fiducial Milky Way composite prior
on the binary 
period distribution to priors that preferentially select short period binary 
solutions. Solutions whose priors prefer short periods (e.g., flat (red 
dashed line) and exponential (blue dot-dashed line) mean period priors) have 
dispersion probability densities that agree remarkably well.  This is true even 
when both the mean period and the width of the period distribution are biased 
low (green dotted line).}
\end{figure*}

\section{Binary population of Segue 1}\label{sec:mean_period_results}

We showed in the previous section that the binary correction to the velocity 
dispersion of Segue 1 is likely to be small, in spite of the large velocity 
variations observed for a few of the stars. In this section, we investigate the 
corresponding constraints on the binary population of Segue 1 obtained by our 
method. By marginalizing over all other parameters, we infer probability 
distributions in the binary fraction $B$, mean log-period $\mu_{\log P}$, and 
width of the period distribution $\sigma_{\log P}$. We find that while the 
binary fraction and width of the period distribution are poorly constrained, 
the mean period is much better constrained. In Figure~\ref{fig:mup} we plot the 
posterior probability distribution of the mean log-period $\mu_{\log p}$. The 
most probable inferred mean period is $\approx$ 10 years. This is significantly 
shorter than the 180 year mean period of solar neighborhood field binaries, 
although a 180 year mean period is still allowed at the $1\sigma$ level.
As shown by the dotted line in Figure~\ref{fig:mup},  if the giants 
are excluded from the sample, the mean period is longer. 
Given the small sample of RGB stars, this is likely because excluding the giants removes the
one star that shows strong evidence of binary orbital motions with a period of $\sim$1 yr.
To see how the inferred intrinsic dispersion is affected by the mean period, in 
Figure~\ref{fig:mupvs}(a) we plot the joint probability distribution of mean 
period and intrinsic dispersion. If the inferred dispersion is larger than 1 
km~s$^{-1}$, the probability distribution of the mean period follows that derived in 
the full sample (see Figure~\ref{fig:mup}). By contrast, the region where the 
inferred dispersion is smaller than 1 km~s$^{-1}$ is dominated by mean periods of 
$\sim$ 2 years---if the mean period were longer, binaries could not possibly 
account for the observed $\approx$ 4 km~s$^{-1}$ dispersion (see the dashed line in 
Figure ~\ref{fig:mup}).  However, the bulk of the probability is in the region 
of $\sigma > $ 1 km~s$^{-1}$, even though the mean period appears to be shorter than 
that of solar neighborhood field binaries.

While the width of the period distribution $\sigma_{\log P}$ is poorly 
constrained on its own, Figure~\ref{fig:sigpvsmup} shows there is a correlation 
between the inferred mean period $\mu_{\log P}$ and $\sigma_{\log P}$ which is 
particularly noticeable for mean periods $\gtrapprox$ 100 years. This reflects 
the fact that for long mean periods, the period distribution must be 
sufficiently wide to include periods short enough to account for the observed 
velocity variations.  More striking is the correlation between the inferred 
binary fraction and the mean period, shown in Figure~\ref{fig:bvsmup}.  This 
relation shows, for example, that if the mean period were shorter than 
$\approx$ 1 month, the binary fraction would have to be smaller than $\approx$ 
0.3; otherwise, the binaries would generate a large non-Gaussian tail in the 
velocity distribution that is inconsistent with the data. This constraint is 
important in that it places an effective upper bound on the probability that 
the observed dispersion of Segue 1 is entirely due to binaries, a bound which 
is almost prior-independent. To show this, in the left panel of 
Figure~\ref{fig:binprior} we plot the probability distribution of the mean 
period using three different priors on the mean period: the Milky Way
composite prior 
discussed in Section \ref{sec:priors}, a flat prior with a minimum mean period 
of one week, and an exponential prior that is strongly biased to short mean 
periods. We find the most probable inferred mean period is only slightly 
different between the flat prior and Milky Way composite prior, while
the exponential  
prior produces very short mean periods of less than one year.  However, as we can 
see in the right panel of Figure~\ref{fig:binprior}, the effect on the inferred 
dispersion is minor. Even though the exponential prior is biased to very short 
mean periods, the binary fraction is constrained to be smaller than 0.2 in that 
case, which limits the effect that such an extreme binary population can have 
on the observed dispersion.  In essence, if a large fraction of the stars were 
short period binaries, then the data would have revealed them.  The data did 
not, which forces the binary fraction to be small, thus limiting the effect on 
the dispersion.

\begin{figure}
\rotatebox{270}{\includegraphics[height=0.964\hsize]{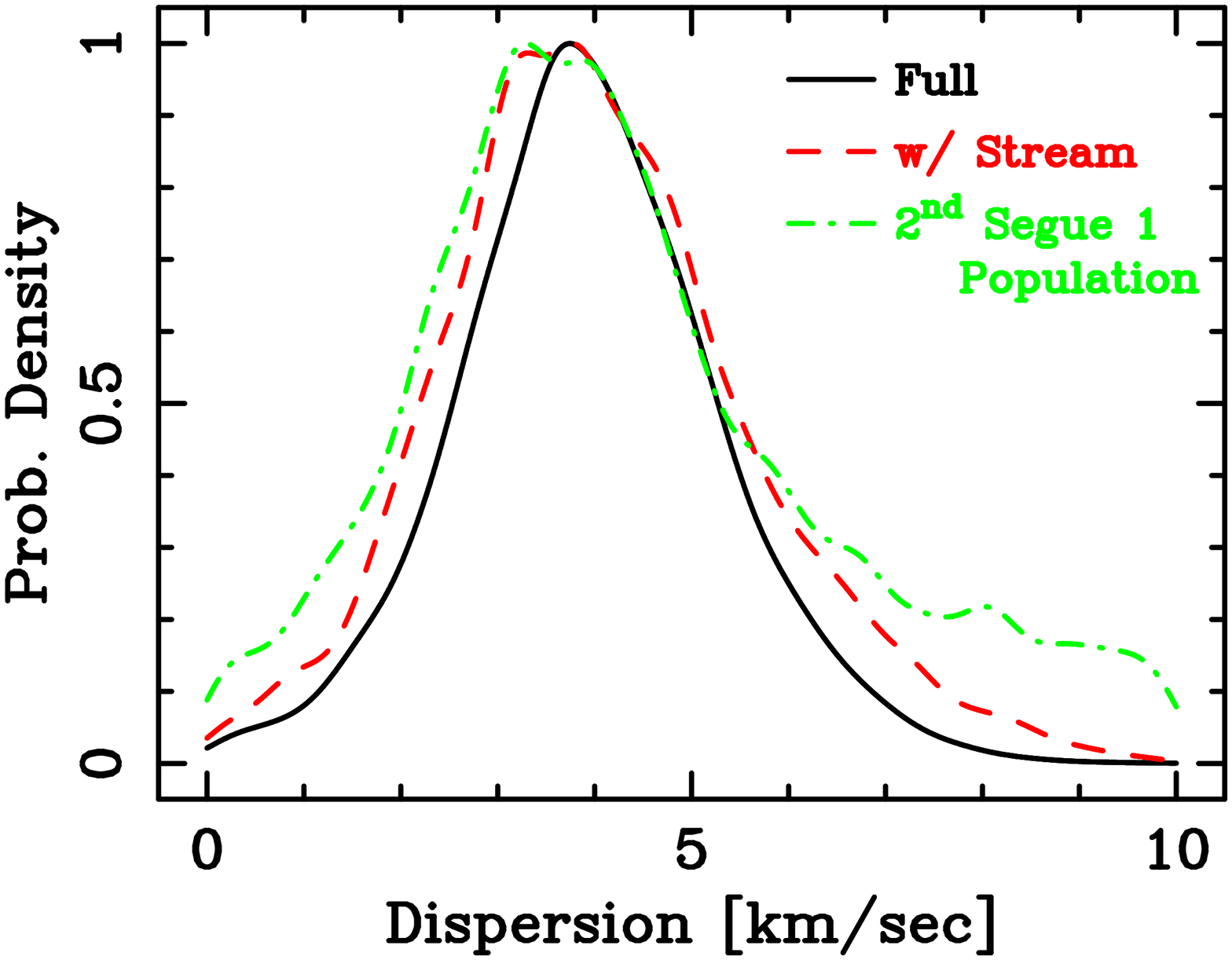}}
\caption{\label{fig:3pop}\footnotesize Plotted curves are the probability 
density of the dispersion assuming the fiducial model with a Milky Way and 
Segue 1 stellar populations (solid black), assuming in addition a third 
stream-like population (dashed red), or assuming a third Segue 1-like 
population (dash-dotted green).  Reassuringly, we find that the extra degrees 
of freedom in terms of the third population did not significantly affect 
the posterior for the intrinsic dispersion.} 
\end{figure}

\section{Contamination by a spatially overlapping tidal
  stream?}\label{sec:stream} 

We can see from Figure \ref{fig:binneddata} that it is unlikely
current data can 
pick out a third population in addition
to the Segue 1 and Milky  Way populations.  To test this, we
considered two cases. For the first case, we assumed that the third
population is spatially uniform, consistent with an overlapping tidal
stream (e.g., Sagittarius stream, see \citealt{niederste2009}).  
The velocity and metallicity distributions of the stream were
  varied independent of the Segue 1 and Milky Way distributions.
For the second case, we 
again assumed the third population has
 its own independent velocity and metallicity distributions, but 
has a spatial distribution given by a modified Plummer profile whose
Plummer radius and outer slope  are allowed to vary independently 
from that of the primary 
Segue 1 population. The second case, therefore, is a test to see if there is
any evidence for two stellar populations in Segue 1. In both cases, we
assumed that the binary period, ellipticity and mass ratio
distributions for the third population were the same as for the
primary Segue 1 population. 
{
Additionally, the priors for parameters defining the
  third population are set to be the same as that for Segue 1 (which
  are listed in Table~\ref{tab:params}).}  
Our results show that
with the current data set, the third population is unconstrained. 
This implies that there is not a third population that is significantly
offset from Segue 1 in its spatial and velocity distribution to be
detected in this data set. Turning this argument around and
discussing how different a third population has to be to be detected
in this data set is an involved question that takes us well beyond the
aims of the present paper. We do, however, show in
Figure~\ref{fig:3pop} that the extra degrees of freedom in terms
of the third population does not significantly affect the inferred
probability distribution of the intrinsic dispersion of Segue 1. 

These tests reinforce the analysis in Paper I where we focused 
specifically on the issue of contamination by Sagittarius stream stars. The 
results of a Monte Carlo analysis there showed that it was unlikely that the 
measured dispersion for Segue 1 was due to a small number of stream stars 
contaminating the sample. The analysis described in this section echoes that 
result and shows in a rigorous statistical manner that there is no evidence for 
a third population given the present sample of velocities. 

\section{Conclusions}\label{sec:conclusion}

We have introduced a comprehensive Bayesian method to analyze multi-epoch 
velocity measurements of Milky Way satellites that incorporates uncertainties 
due to imperfect knowledge of membership and binary orbital motion of stars.  
We applied this method to Segue 1 using the kinematic data set described in 
Paper I, which includes 181 candidate member stars, 67 of which have repeat 
measurements.  We model the likelihoods of relevant populations (Milky Way, 
Segue 1, and possibly an overlapping stream) and thereby incorporate membership 
probabilities implicitly in the method.  Our results support the interpretation 
that Segue 1 is a dark-matter-dominated galaxy with an intrinsic velocity 
dispersion of 3.7$^{+1.4}_{-1.1}$ km~s$^{-1}$ at $1\sigma$. We stress here that the
multi-epoch data analysis is critical---with just the average
velocity for each star, the possibility that most of the observed
dispersion of Segue 1 is due binary orbital motion cannot be
disfavored.  

Our method produces a posterior for the membership probability of each star and 
simultaneously constrains the radial distribution of Segue 1 member stars 
without appealing to separate photometry.  Using the full likelihood
(see Equation~(\ref{eq:lgalmw}) and Table~\ref{tab:params}), we find $R_{1/2}
= 28^{+5}_{-4}$,  which is in excellent agreement with past
photometric measurements  \citep{martin2008}. We also included the 
slope of the stellar profile (see Equation~(\ref{eq:modplum})) in our full
likelihood analysis and find that $\alpha=4.1^{+2.0}_{-0.8}$, which is
consistent with the standard Plummer profile ($\alpha=5$). 

To include the possibility that each star is in a binary system, we modified 
the velocity likelihood of Segue 1 to take into account changes in the velocity 
distribution resulting from binary orbital motion. The binary properties of the Segue 1 
ensemble were parameterized by a log-normal distribution in period and a total 
binary fraction. Only the mean period was marginally constrained, with a most 
probable mean period of 10 years, much smaller than the 180 year mean 
period for binary stars in the solar neighborhood. However, our results are 
still consistent with a mean period of 180 years at about $1\sigma$.

We also found a slight degeneracy between the binary fraction and mean period 
with the binary fraction decreasing with the mean period. 
The case where a large fraction of Segue 1 stars are short period binaries 
($P \lesssim 1$ yr) is disfavored by the lack of large velocity variations 
(relative to the errors) in the repeat measurements.
One implication is that our inferred  intrinsic velocity 
dispersion is robust to the period distribution. We explicitly tested this by 
varying the priors on the binary parameters and found no significant affect on the inferred intrinsic 
dispersion probability distribution. The inferred intrinsic dispersion was also
insensitive to the inclusion or exclusion of velocity outliers. 

Our results show that the velocity dispersion of Segue 1 is larger than 1 km~s$^{-1}$ 
at the 98.3\% confidence level. The small probability of dispersions lower 
than 1 km~s$^{-1}$ is caused by the possibility of binary stars with short periods 
inflating the velocity dispersion.  Note that with a 1 km~s$^{-1}$ velocity
dispersion, Segue 1 would have $(M_{1/2}/L_{1/2}) \simeq 150$ within the half-light
radius (if interpreted as a system in equilibrium) and therefore 
would still be among
the most dark-matter-dominated satellites 
of the Milky Way \citep{walker2009a,wolf2010}. An alternative
interpretation at this  confidence level would be that Segue 1 is a
star cluster that is disrupting and hence had its intrinsic velocity
dispersion inflated to about 1 km~s$^{-1}$, and with parameters such that
binary orbital motion contributes an additional $\sim$3 km~s$^{-1}$
dispersion.  Beyond the low probability we determined, this 
possibility seems unlikely on other grounds as well.
The Jacobi radius at a distance of 23 kpc with 
$(M/L)_{\rm stellar} \sim 10$ (an extreme value) is
about 30 pc---smaller than the 
region covered by our Segue 1 sample of stars (about 70 pc).   Hence,
we expect to see tidal features \citep{penarrubia2009} and none could
be identified (see Paper I for more details).

From the inferred probability distribution for the intrinsic
dispersion, we find that there is only about 0.4\% probability that
the intrinsic dispersion is smaller than about 0.3 km~s$^{-1}$. 
Interpreted as an equilibrium system, such a dispersion would imply $(M_{1/2}/L_{1/2})_{\rm
  stellar} < 10$. At this confidence level, therefore, the stellar
velocity data  allow for the possibility that Segue 1 is a star
cluster, albeit with a rather extreme stellar population, and with the
measured velocity dispersion dominated by the orbital motion of binary
stars with mean periods of around a year (see 
Figure~\ref{fig:mup}). However, given the tidal arguments above,
it is unclear how we may think of this system as being in equilibrium
when it is not dark matter dominated.  In addition, the large measured 
metallicity spread is also not consistent with the star cluster hypothesis (Paper I).

The most likely interpretation of our results is that Segue 1 is a dark 
matter dominated galaxy.  In this case, our inferred velocity dispersion
implies a mass of 
$M_{1/2} = 5.8^{+8.2}_{-3.1} \times 10^5 M_\odot$ 
within a sphere that encloses half the galaxy's stellar luminosity,
which from our full likelihood analysis is $r_{1/2} = 36^{+8}_{-5}$
pc. To calculate this mass, we have used $M_{1/2}  = 3 G^{-1} r_{1/2}
\sigma^2$ from \cite{wolf2010} along with 
the distribution for $r_{1/2}$ and $\sigma$ that we have derived using
the full likelihood. 
The average density of dark matter within this radius is therefore  
$\bar{\rho}_{1/2} = 2.5^{+4.1}_{-1.9} M_\odot {\rm pc}^{-3}$,
which is the highest density of dark matter yet measured in any Local
Group object \citep{wolf2010,walker2009a}.  
We note here that
a flat prior in the intrinsic dispersion has been used throughout the paper.
If a flat prior in $M_{1/2}$ is imposed, the resultant confidence interval changes slightly
to $M_{1/2} = 5.2^{+8.5}_{-3.8} \times 10^5 M_\odot$.

It is worth emphasizing that this dark matter density is among the
highest dark matter densities that is known definitively in any
galaxy. Some of the larger elliptical galaxies are measured to have
comparably high mass densities at their half-light radii ($\sim 1 {\rm
  M}_\odot {\rm pc}^{-3}$), but inferring similarly high dark matter  
density is complicated by the fact that galaxies of this type are
baryon-dominated in their centers
\citep[e.g.][]{cappellari2006,tollerud2010}.    For rotation-supported
galaxies, the highest 
dark matter density would be obtained at the innermost point where
rotation velocity is reliably measured. For example, some of the
nearby low surface brightness galaxies in \citet{Kuzio08}
show rotation speeds of 10--40 km~s$^{-1}$ at about 300 pc, or average dark
matter density in the range 0.05--1 ${\rm M}_\odot/{\rm pc}^3$.  
For reference if we assume a typical NFW profile for the Milky Way,
then the Milky Way will have this density in dark matter at a radius
of about 100 pc (assuming cold dark matter model). At the other end of
the mass range for a halo with $V_{\rm max}$ less than 10 km~s$^{-1}$, these
densities will occur at radii smaller than the half light radius of
Segue 1 (assuming cold dark matter), suggesting that Segue 1 has a
more massive halo.  A third way of inferring the total mass in a
galaxy is through strong lensing, which provides a measurement of the
surface mass density. In order for  strong lensing to occur, the
surface mass density has to be larger than a critical value. Taking
the angular diameter distance to the lens, $D=1$ Gpc, assuming that
the lens is halfway between the source and observer, and using $\sim$
arcsec for the angular size of  a typical Einstein ring
\citep{bolton2008} yields a characteristic total mass density (not all
dark matter) within the Einstein ring of 
$c^2/(2\pi G D^2 {\rm   arcsec})$ $\sim$ $1 M_\odot/{\rm pc}^3$.      

If we assume that all the dwarf spheroidal satellite galaxies of the
Milky Way inhabit similar halos \citep{strigari2008}, this high density
implies that Segue 1 has the highest phase space density among all
dwarfs and hence should provide the best constraints on thermal and
non-thermal warm dark matter. The large determined dark matter halo
mass also validates previous expectations
\citep{geha2009,martinez2009} that Segue 1 is an excellent target for
the indirect detection of dark matter and a useful laboratory for
studying galaxy formation at the extreme faint end of the luminosity
function. 

\section*{Acknowledgments}

G.D.M. gratefully acknowledges generous support by Gary McCue. This
work was partially supported at UCI by NSF grant PHY--0855462 and NASA
grant NNX09AD09G. 

\bibliographystyle{apj}
\bibliography{segue}

\end{document}